\documentclass[a4paper]{article}
\pdfoutput=1

\usepackage{macros}

\preprint{ }
\title{Central Charges and Vacuum Moduli of 2d $\cN=(0,4)$ Theories from Class $\cS$}

\author[1]{Wei Cui,}
\author[2]{Junkang Huang,}
\author[2]{Zi-Xiao Huang,}
\author[2]{Satoshi Nawata,}
\author[2]{Shutong Zhuang}

\affiliation[1]{Beijing Institute of Mathematical Sciences and Applications (BIMSA)
Huaibei Town, Huairou District, Beijing 101408, China}
\affiliation[2]{Department of Physics and Center for Field Theory \& Particle Physics, Fudan University, 20005, Songhu Road, 200438 Shanghai, China}

\emailAdd{cwei@bimsa.cn}
\emailAdd{junkang-huang@outlook.com}
\emailAdd{huangzx22@m.fudan.edu.cn}
\emailAdd{snawata@gmail.com}
\emailAdd{22110190097@m.fudan.edu.cn}

\abstract{We investigate 2d $\mathcal{N}=(0,4)$ supersymmetric theories obtained from a topologically-twisted reduction of 4d $\mathcal{N}=2$ class $\mathcal{S}$ theories on a Riemann surface. This study addresses subtle aspects of central charges, unbroken gauge groups, and emergent superconformal R-symmetries of these theories. Focusing on infrared vacuum structures, we propose conjectural formulas for the central charges. For theories with the gauge group $SU(2)$, we use Lagrangian descriptions to analyze the vacuum moduli spaces. In particular, we examine two distinct branches—the special Higgs branch and the twisted Higgs branch—by computing their Hilbert series, and find agreement with the proposed central charge formulas.}

\def\ii{\mathrm{i}}
\newcommand{\ap}{a^{\prime}}
\newcommand{\bp}{b^{\prime}}
\renewcommand{\cp}{c^{\prime}}

\def\Dcl{\mathcal{D}}

\begin{document}
\allowdisplaybreaks

\maketitle

\section{Introduction}\label{sec:introduction}

The study of quantum field theory (QFT) has repeatedly uncovered intriguing and often unexpected relationships among different theories. One of the most powerful frameworks for revealing and exploiting such relations is dimensional reduction. By compactifying a $d$-dimensional QFT on an $n$-dimensional compact manifold, one obtains an effective QFT in $d-n$ dimensions, whose dynamics are governed by both the original theory and the geometry of the compactification manifold. 

A central question in this context concerns the relationship between the higher- and lower-dimensional QFTs. 
In particular, one would like to understand which structural features of the original theory survive the dimensional reduction, how they are encoded in the effective lower-dimensional theory, and in what ways genuinely new phenomena may emerge. Among the various aspects of this relationship, it is important to understand the behavior of symmetries under dimensional reduction.
Some symmetries of the higher-dimensional theory may be inherited, possibly in a reduced or modified form, while others may be explicitly broken by the geometry of the compactification manifold or by the choice of background fields. Conversely, new symmetries may emerge dynamically in the infrared of the lower-dimensional theory, with no manifest realization in the ultraviolet description. Understanding the fate of symmetries under dimensional reduction, and distinguishing between inherited, broken, and emergent symmetries, is a key step toward elucidating the structure of QFTs across dimensions.

From this perspective, the 6d $\mathcal{N}=(2,0)$ superconformal field theory (SCFT), which describes the effective worldvolume theory on M5-branes,  occupies a distinctive position since it represents the maximally supersymmetric conformal theory in the highest possible spacetime dimension~\cite{Nahm:1977tg}. Despite the absence of a Lagrangian description, it governs a remarkably rich structure that can be probed indirectly through its compactifications, symmetries, protected observables, and associated moduli spaces.

Compactifying the 6d $\cN=(2,0)$ SCFT of ADE type on various manifolds with an appropriate topological twist, we obtain a plethora of lower-dimensional supersymmetric QFTs. The resulting theories, typically denoted $\mathcal{T}_G[M]$ for a compactification manifold $M$, not only provide explicit constructions of supersymmetric QFTs in various dimensions but also reveal intriguing relations between the supersymmetric theory $\mathcal{T}_G[M]$ and the geometry and topology of $M$. 
 
A celebrated example arises from compactifying the 6d $\mathcal{N}=(2,0)$ SCFT on a punctured Riemann surface $C_{g_1,n}$, generating a large class of 4d $\mathcal{N}=2$ SCFTs known as class $\mathcal{S}$ theories~\cite{Gaiotto:2009we,Gaiotto:2009hg}.
Over the years, the study of class $\cS$ theories has led to remarkable developments and produced profound results in understanding the physical and mathematical structures of 4d $\cN=2$ theories. 

Given these developments, a natural next step is to reduce a class $\cS$ theory $\mathcal{T}_G[C_{g_1,n}]$ further on a Riemann surface $C_{g_2}$. By performing a topological twist in which the $U(1)_{C_{g_2}}$ holonomy of $C_{g_2}$ is embedded into the $U(1)_r$ subgroup of the 4d R-symmetry $SU(2)_R \times U(1)_r$, one obtains a 2d $\mathcal{N}=(0,4)$ theory~\cite{Bershadsky:1995vm,Kapustin:2006hi,Benini:2013cda}.

The special case $C_{g_2}=S^2$ has been studied in~\cite{Putrov:2015jpa,Nawata:2023aoq}. In this situation, the resulting $(0,4)$ theory flows to a non-linear sigma model (NLSM) on the hypermultiplet moduli space (the Higgs branch), equipped with a non-trivial left-moving bundle. For genus $g_1>0$, the Cartan subgroup of the gauge group is unbroken at a generic point of the Higgs branch, so a $U(1)^{r_{G}g_1}$ gauge symmetry survives~\cite{Hanany:2010qu}, where $r_G$ is the rank of $G$. In two dimensions, such unbroken Abelian gauge sectors are gapped in the infrared~\cite{Witten:1997yu}.
This phenomenon leads to subtle infrared structures and renders the computation of central charges non-trivial.

In this paper, we generalize this setup and undertake a systematic analysis of the infrared structure of the 2d $\mathcal{N}=(0,4)$ theories obtained from the dimensional reduction of $\mathcal{T}_G[C_{g_1,n}]$ on a general Riemann surface $C_{g_2}$. A central question we address is the relationship between the symmetries of the 4d $\cN=2$ parent SCFT and 2d $\cN=(0,4)$ theory.
In many examples of dimensional reduction, 't Hooft anomaly coefficients of the lower-dimensional theory can be obtained by integrating the higher-dimensional anomaly polynomial over the compactification manifold~\cite{Benini:2009mz,Tachikawa:2015bga}. However, in the present setting, the information relevant for determining central charges can be lost in this integration process. As a result, the anomaly polynomial obtained after integration does not necessarily encode information about the infrared central charges.

These difficulties share a common origin: both the anomaly polynomial integration and naive 't~Hooft matching rely on UV data that fails to capture the emergent IR R-symmetry. When an infrared-gapped Abelian gauge sector is present, the identification of the infrared R-symmetry from the UV ones, and hence the application of 't~Hooft anomaly matching to the right-moving central charge, fails to produce the correct result. We therefore turn to a direct analysis of the infrared physics of the resulting 2d $(0,4)$ theories, from which we propose a set of conjectural formulas for their infrared central charges.

To determine the central charges, we study the structure of the vacuum moduli space for this class of 2d $\mathcal{N}=(0,4)$ theories. While the vacuum moduli space generically has several branches (irreducible components), we concentrate on two of them: the special Higgs branch and the twisted Higgs branch. The \textit{twisted Higgs branch} is parametrized solely by vacuum expectation values (VEVs) of twisted hypermultiplets. The \textit{special Higgs branch} is characterized by the maximum number of hypermultiplet scalar VEVs, whereas it is generally supported by simultaneous VEVs of both hypermultiplets and twisted hypermultiplets. Each branch carries its own affine $SU(2)$ R-symmetry current, reflecting distinct infrared realizations of the small $(0,4)$ superconformal algebra. In particular, on branches where an unbroken Abelian gauge sector is present in the ultraviolet, the superconformal $SU(2)$ R-symmetry is emergent in the infrared.  

To make the infrared physics more explicit, we analyze these moduli spaces for the gauge group $G={SU}(2)$ where an explicit Lagrangian description is available. By computing the Hilbert series of the special Higgs branches and the twisted Higgs branches, we elucidate detailed features of the vacuum moduli space and provide explicit checks of our proposed formulas for the central charges.

The paper is organized as follows. In Section~\ref{sec:M5-reduction}, we study 2d $\mathcal{N}=(0,4)$ theories obtained from the dimensional reduction of 4d $\mathcal{N}=2$ class $\mathcal{S}$ theories on Riemann surfaces, with a focus on determining their infrared central charges. After briefly reviewing the basic features of class $\mathcal{S}$ theories and their reduction to two dimensions, we discuss the structure of $(0,4)$ multiplets, R-symmetries, and anomalies. We then explain the issues of naive approaches to central charge computations based on ’t~Hooft anomaly matching and the integration of the anomaly polynomial of the higher-dimensional parent theory. Motivated by these observations, we formulate conjectures for the infrared central charges of the resulting $(0,4)$ theories by directly analyzing the massless spectrum after the Higgs mechanism.

Section~\ref{sec:vac-moduli} is devoted to a detailed study of the vacuum moduli spaces for the case $G = SU(2)$, where an explicit $(0,4)$ Lagrangian description is available. We first analyze the vacuum moduli equations and then study in detail the special Higgs branch and the twisted Higgs branch using Hilbert series techniques. This analysis not only provides explicit checks of the conjectured central charges but also reveals the emergence of a superconformal $SU(2)$ R-symmetry in the infrared.

Several technical details and complementary discussions are collected in the appendices. Appendix~\ref{app:notations} summarizes our notational conventions and provides a unified reference for symbols, representations, and normalization choices used throughout the paper. Appendix~\ref{app:Higgs} contains a detailed analysis of the Higgs mechanism for the 2d (0,4) theories with gauge group $SU(2)$, which allows us to read off the massless spectrum after Higgsing. In Appendix~\ref{app:hilbert-series}, we present explicit examples of Hilbert series computations for the special Higgs branch, illustrating the general methods employed in the main text and providing concrete examples. Appendix~\ref{app:single_M5} reviews the dimensional reduction of a single M5-brane, providing a simple and instructive check of our conjectures.

\section{2d \texorpdfstring{$\mathcal{N}=(0,4)$}{N=(0,4)} theory from class \texorpdfstring{$\cS$}{S} on Riemann surfaces}\label{sec:M5-reduction}

In this section, we study 2d $(0,4)$ supersymmetric theories obtained from the dimensional reduction of the class~$\mathcal{S}$ theory $\mathcal{T}_G[C_{g_1,n}]$ on a Riemann surface $C_{g_2}$. After briefly reviewing the basic features of class~$\mathcal{S}$ theories, we describe the construction of the resulting $(0,4)$ theory by clarifying the structure of $(0,4)$ multiplets and the associated R-symmetries.

We then turn to the computation of the central charges of the resulting 2d $(0,4)$ infrared CFTs. In particular, we analyze the limitations of both the naive application of ’t~Hooft anomaly matching and of the standard approach based on integrating the higher-dimensional anomaly polynomial. These limitations are attributed to the presence of unbroken Abelian gauge groups and to the identification of the superconformal $SU(2)$ R-symmetry in the infrared. By understanding the origin of this mismatch, we are led to propose conjectural formulas for the infrared central charges of the $(0,4)$ theories, by counting the massless spectrum after the Higgs mechanism.

\subsection{Class \texorpdfstring{$\cS$}{S} theories and reduction to 2d \texorpdfstring{$\mathcal{N}=(0,4)$}{N=(0,4)} theories}\label{sec:4dto2d}

Class $\cS$ theories are 4d $\cN=2$ theories that arise by compactifying the 6d
$\mathcal{N}=(2,0)$ theory of type $G$ on a punctured Riemann surface
$C_{g_1,n}$ of genus $g_1$ with $n$ punctures~\cite{Gaiotto:2009we,Gaiotto:2009hg}.
Consider the twisted compactification of 6d $\cN=(2,0)$ SCFT $\cT_G$ on a Riemann surface $C_{g_1,n}$ of genus $g_1$ with $n$ punctures.  
To preserve 4d $\cN=2$ supersymmetry, one turns on a partial topological twist that embeds the $U(1)_{C_{g_1,n}}$ holonomy of $C_{g_1,n}$ into the $SO(5)_R$ symmetry.  
After the twist, the resulting 4d theory depends only on the complex structure of $C_{g_1,n}$ and the local data specified at its punctures.  
We denote this theory by $\cT_G[C_{g_1,n}]$,
and refer to it as a \emph{class~$\cS$ theory} of type~$G$.\footnote{More precisely, extended operators in class~$\cS$ theories are sensitive to the global structure of the gauge group. For this, one must specify additional data in the form of a maximal isotropic sublattice of $H^{1}(C_{g_1,n}, Z(G))$~\cite{Aharony:2013hda}. We will not pursue these subtleties in this paper.}

In this paper, we consider only regular punctures, which are characterized by embeddings 
\be 
  \rho: \mathfrak{su}(2) \hookrightarrow \mathfrak{g}~,
\ee 
or equivalently by nilpotent orbits in $\mathfrak{g}$~\cite{Chacaltana:2012zy}.  
They determine local boundary conditions for fields near each puncture and give rise to flavor symmetries in the resulting 4d theory. The resulting 4d effective theory is an $\cN=2$ SCFT with $SU(2)_R\times U(1)_r$ R-symmetry, and it does not generically admit a Lagrangian description. 

Since any Riemann surface can be decomposed into three-punctured spheres, the theory $\cT_G[C_{g_1,n}]$ may be assembled by gauging together copies of the \emph{trinion} theories $\cT_G[C_{0,3}]$, which serve as fundamental building blocks.
The complexified gauge couplings of $\cT_G[C_{g_1,n}]$ are identified with the complex structure moduli of the curve $C_{g_1,n}$, and different pair-of-pants decompositions of $C_{g_1,n}$ correspond to different weakly coupled frames of the same 4d theory.
Changing the pants decomposition amounts to an action of the mapping class group, which manifests in the 4d theory as generalized S-duality~\cite{Gaiotto:2009we}.

For a class~$\cS$ theory $\cT_G[C_{g_1,n}]$ with all maximal punctures, the effective
numbers of 4d vector and hypermultiplets can be expressed in terms of
group-theoretic data~\cite{Tachikawa:2015bga}:
\begin{equation}
\label{eq:nvnh-general}
n_v=\left(\frac{2}{3} h^\vee_G d_G+\frac{r_G}{2}\right)(2g_1 - 2 + n)
-\frac12\, d_G\, n~,
\qquad
n_h=\frac{2}{3}\, h^\vee_G d_G\,(2g_1 - 2 + n)~,
\end{equation}
where $h^\vee_G$, $r_G$, and $d_G$ denote the dual Coxeter number, rank,
and dimension of the Lie algebra of $G$, respectively. For $G=SU(N)$, $h^\vee_G=N,~r_G=N-1,~d_G=N^2-1$, the results are
\begin{equation}\label{nvnh_general_punc}
\begin{aligned}
    n_v&=\frac{1}{3}(g_1-1)(N-1)\left[4N(N+1)+3\right]+\frac{1}{6}N(N-1)\left(4N+1\right)n~,\\
    n_h&=\frac{2}{3}N(N^2-1)\left[2(g_1-1)+n\right]~.
    \end{aligned}
\end{equation}

\begin{figure}
    \centering
    \includegraphics[width=0.5\linewidth]{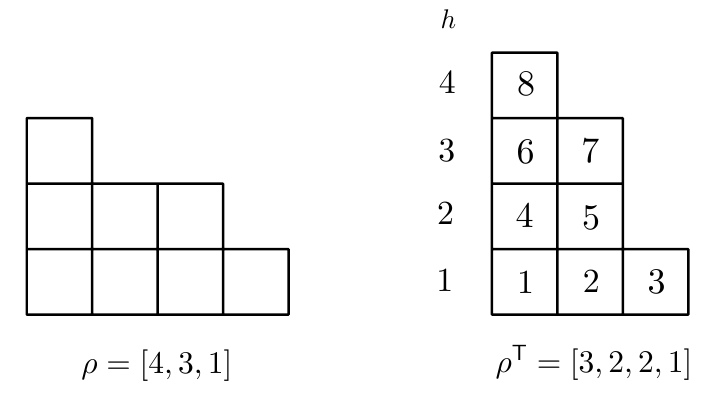}
    \caption{For a puncture labeled by $\rho=[4,3,1]$ when $N=8$, label the boxes in $\rho^{\mathsf{T}}$ by $k=1,\cdots,8$ as on the right, and the corresponding $h_k$ are $(1,1,1,2,2,3,3,4)$, and consequently $n_v(\rho)=199,~ n_h(\rho)=204$.}
    \label{fig:young-diag}
\end{figure}

Generally~\cite{Chacaltana:2012zy,Tachikawa:2015bga}, for $G=SU(N)$ with general punctures labeled by partitions $\rho^{(i)}, i=1,\cdots,n$ of $N$:
\begin{equation}
    n_v=(g_1-1)\left(\frac{4}{3}h^\vee_G d_G+r_G\right)+\sum_{i=1}^nn_v\left(\rho^{(i)}\right)~,\quad n_h=(g_1-1)\frac{4}{3}h^\vee_G d_G+\sum_{i=1}^nn_h\left(\rho^{(i)}\right)~,
\end{equation}
where $n_v(\rho)$ and $n_h(\rho)$ are given by
\begin{equation}
    n_v(\rho)=\sum_{k=1}^{N}(2k-1)\left(k-h_k(\rho^{\mathsf{T}})\right)~,\quad n_h(\rho)=n_v(\rho)+\frac{1}{2}\left(\sum_{a=1}^{|\rho^{\mathsf{T}}|}(\rho^{\mathsf{T}}_{a})^2-N\right)~.
\end{equation}
where $h_k(\rho^{\mathsf{T}})$ is the row of the $k$-th box in the Young diagram associated with $\rho^{\mathsf{T}}$, with rows sorted in non-decreasing order. For example, as in Figure \ref{fig:young-diag}, when $\rho=[4,3,1]$, $\rho^{\mathsf{T}}=[3,2,2,1]$ and $h_k=(1,1,1,2,2,3,3,4)$, one has $n_v(\rho)=199,~ n_h(\rho)=204$.

As another example, for a full puncture $\rho=[1^N]$, $\rho^{\mathsf{T}}=[N]$, and $h_k([N])=1$ for all $k$, 
\begin{equation}
    n_v([1^N])=\frac{1}{6}N(N-1)(4N+1)~,\quad n_h([1^N])=\frac{2}{3}N(N^2-1)~,
\end{equation}
the total numbers of multiplets are therefore given by \eqref{nvnh_general_punc}.

We now consider the dimensional reduction of the 4d $\mathcal{N}=2$ class~$\cS$ theory on $C_{g_2}$.
The holonomy $U(1)_{C_{g_2}}$ of $C_{g_2}$ breaks supersymmetry unless we perform a topological twist.
To obtain a 2d theory with $\cN=(0,4)$ supersymmetry, we perform the topological twist by taking the diagonal subgroup of the $U(1)_{C_{g_2}}$ holonomy group and $U(1)_r$:
\begin{equation} \label{eq:U1r_twist}
U(1)^{\mathrm{tw}}_{C_{g_2}}
\;=\;
\mathrm{diag}\left( U(1)_{C_{g_2}} \times U(1)_r \right)~.
\end{equation}
To see this explicitly, recall that under
\be 
SO(4) \;\to\; SO(2) \times U(1)_{C_{g_2}}~,
\ee 
the 4d supercharges split into left- and right-moving 2d spinors,
carrying $U(1)_{C_{g_2}}$ charge $q_2=\pm \frac12$ and $U(1)_r$ charge $q_r=\pm \frac12$.
Precisely those supercharges with
$
q_{2}+q_r=0
$
become scalars on $C_{g_2}$ and remain unbroken.
This leaves four right-moving supercharges and no left-moving ones,
so the resulting 2d theory has $\mathcal{N}=(0,4)$ supersymmetry~\cite{Putrov:2015jpa,Gadde:2015wta}.

\subsection{(0,4) multiplets, R-symmetries, and anomalies}
\label{subsec:2d-fields}
The structure of fields and symmetries in 2d $(0,4)$ theories has been systematically analyzed in~\cite{Tong:2014yna}.  
A detailed derivation of the $(0,4)$ reduction of 4d $\cN=2$ theories on a Riemann surface $C_{g_2}$ is provided in the appendices of~\cite{Putrov:2015jpa,Cecotti:2015lab}.  
To be self-contained, we briefly summarize the relevant results here.

\begin{figure}[ht]
    \centering
    \includegraphics[width=0.5\linewidth]{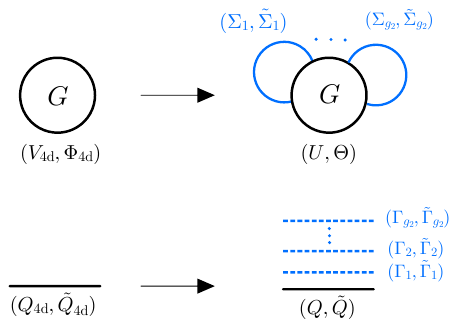}
    \caption{The $\mathcal{N}=(0,4)$ reduction of the 4d $\mathcal{N}=2$ multiplets on a Riemann surface of genus $g_2$. For simplicity, in the remaining part, we will use a single solid blue line to represent $g_2$ twisted hypermultiplets $(\Sigma_j,\tilde\Sigma_j)$, and a single dashed blue line to represent $g_2$ Fermi multiplets $\Gamma_j$.}
    \label{fig:reduce_fields}
\end{figure}

The R-symmetry of a 2d $\cN=(0,4)$ theory is $SO(4)_R \;\cong\; SU(2)_- \times SU(2)_+$,
and the R-charges of the various $(0,4)$ multiplets are listed in Table~\ref{tab:(0,4)-fields}. 
When the parent 4d theory admits a Lagrangian description, its dimensional reduction on $C_{g_2}$ produces the UV field content in two dimensions~\cite{Putrov:2015jpa,Cecotti:2015lab}. The resulting 2d field contents are inherited from the zero modes of 4d multiplets and are organized as follows (also see Figure~\ref{fig:reduce_fields}):
\begin{itemize}
    \item 1 4d vector $(U_{\mathrm{4d}},\Phi_{\mathrm{4d}})$ $\longrightarrow$ 1 $(0,4)$ vector $(U,\Theta)$ + $g_2$ $(0,4)$ twisted hyper $(\Sigma_j,\tilde\Sigma_j)$
    \item 1 4d hyper $(Q_{\mathrm{4d}},\tilde Q_{\mathrm{4d}})$ $\longrightarrow$ 1 $(0,4)$ hyper $(Q,\tilde Q)$ + $g_2$ $(0,4)$ Fermi $(\Gamma_j,\tilde\Gamma_j)$
\end{itemize}
where the index $j=1,\ldots,g_2$ labels the $g_2$ copies originating from holomorphic one-forms $H^0({C}_{g_2},K)$ (with the canonical bundle $K$).
The charges carried by 2d fields are summarized in Table~\ref{tab:UV-charges}.
Based on these charge assignments, one can identify the two R-symmetry factors $SU(2)_{-}$ and $SU(2)_{+}$ of the 2d theory with the symmetries of the parent 4d $\cN=2$ theory as
\begin{equation}
\label{eq:2d-R-identification}
U(1)_+ = U(1)^{\text{tw}^{\prime}}_{C_{g_2}}=\mathrm{diag} \left(U(1)_{C_{g_2}}  \times U(1)^{-1}_r \right)~, 
\qquad
SU(2)_- = SU(2)_R~,
\end{equation}
where $U(1)_+$ is the Cartan subgroup of $SU(2)_+$.

\begin{table}[ht]
\centering
\renewcommand{\arraystretch}{1.3}
\begin{tabular}{c|c|c|c}
\hline
{(0,4) multiplets} & {(0,2) multiplets}   & {Components}          & ${SU(2)_{-} \times SU(2)_{+}}$                            \\ \hline
vector                                  & vector $U=(A_{\mu},\lambda_{-})$ + Fermi $\Theta=(\tilde{\lambda}_-)$               & $A_{\mu},\lambda^a_-$   &   $(\mathbf{1},\mathbf{1}), (\mathbf{2},\mathbf{2})$                            \\
twisted hyper                           & chiral $\Sigma=(\sigma,\lambda_{+}^{\dagger})$ + chiral $\tilde{\Sigma}=(\tilde\sigma,\tilde{\lambda}_+)$ & $\sigma_a,\lambda^b_+$  &  $ (\mathbf{1},\mathbf{2}), (\mathbf{2},\mathbf{1})$  \\
hypermultiplet                          & chiral $Q=(q, \psi_{+})$ + chiral $ \tilde{Q}=(\tilde{q},\tilde{\psi}_+)$          & $q^a,\psi_{+,b}$     & $(\mathbf{2},\mathbf{1}), (\mathbf{1},\mathbf{2})$                   \\
Fermi                                   & Fermi $\Gamma=(\psi_{-})$ + Fermi $\tilde{\Gamma}=(\tilde{\psi}_-)$   & $\psi_-^a$    & $(\mathbf{1},\mathbf{1})$                              \\ \hline
\end{tabular}
\caption{Field contents of 2d $(0,4)$ multiplets expressed in terms of $(0,2)$ multiplets, and their charges under $SU(2)_{-} \times SU(2)_{+}$ R-symmetry. 
}
\label{tab:(0,4)-fields}
\end{table}

\begin{table}[ht]
\centering
\normalsize 
\renewcommand{\arraystretch}{1.3} 
\begin{tabular}{l|ccc|cc}
\hline
Superfield & $U(1)_{R}$ & $U(1)_{r}$ & $U(1)_{C_{g_2}}$& $U(1)_{C_{g_2}}^{\text{tw}}$ & $U(1)^{\text{tw}^{\prime}}_{C_{g_2}}$\\
\hline
$U = (A_{\mu}, \lambda_{-})$ & $\left(0,\frac{1}{2}\right)$ & $\left(0,\frac{1}{2}\right)$ & $\left(0,-\frac{1}{2}\right)$  & $(0,0)$  & $\left(0,-\frac{1}{2}\right)$ \\
$\Theta=(\tilde{\lambda}_{-})$ & $-\frac{1}{2}$ & $\frac{1}{2}$ &$-\frac{1}{2}$ & 0 & $-\frac{1}{2}$ \\
$\Sigma=(\sigma,\lambda_{+}^{\dagger})$ & $\left(0,-\frac{1}{2}\right)$ & $\left(0,-\frac{1}{2}\right)$ & $\left(1,-\frac{1}{2}\right)$ & $(\frac{1}{2},-\frac{1}{2})$  & $\left(\frac{1}{2},0\right)$ \\
$\tilde{\Sigma}=(\tilde{\sigma},\tilde{\lambda}_{+})$& $\left(0,-\frac{1}{2}\right)$& $\left(1,\frac{1}{2}\right)$& $\left(0,\frac{1}{2}\right)$& $(\frac{1}{2},\frac{1}{2})$& $\left(-\frac{1}{2},0\right)$ \\
\hline
$Q=(q,\psi_{+})$ & $\left(\frac{1}{2},0\right)$ & $\left(0,-\frac{1}{2}\right)$ & $\left(0,\frac{1}{2}\right)$ & $(0,0)$  & $\left(0,\frac{1}{2}\right)$ \\
$\tilde{Q}=(\tilde{q}$,$\tilde{\psi}_{+})$ & $\left(\frac{1}{2},0\right)$ & $\left(0,-\frac{1}{2}\right)$ & $\left(0,\frac{1}{2}\right)$& $(0,0)$ & $\left(0,\frac{1}{2}\right)$ \\
$\Gamma=(\psi_{-})$ & 0 & $-\frac{1}{2}$ &$-\frac{1}{2}$ & $-\frac{1}{2}$  & 0\\
$\tilde{\Gamma}=(\tilde{\psi}_{-})$ & $0$ & $-\frac{1}{2}$ &$-\frac{1}{2}$&  $-\frac{1}{2}$ & 0 \\
\bottomrule
\end{tabular}
\caption{Charges of the 2d $\mathcal{N}=(0,4)$ superfields obtained by the dimensional reduction of a 4d $\mathcal{N}=2$ theory on $C_{g_2}$.  
For each superfield, we list its charges under the Cartan subgroup $U(1)_R \subset SU(2)_R$ of the 4d $SU(2)_R$ symmetry, the twisted Lorentz symmetry $U(1)^{\mathrm{tw}}_{C_{g_2}}$, and the 4d $U(1)_r$ symmetry.  
These assignments follow from the $U(1)_r$ twist on $C_{g_2}$ and determine the resulting $(0,4)$ multiplet structure in two dimensions.}
\label{tab:UV-charges}
\end{table}

\paragraph{Remark.} Throughout this paper we study the 2d $\cN=(0,4)$ theory obtained via
\emph{dimensional reduction} of class $\cS$ theories on $C_{g_2}$,
retaining only the zero modes and discarding all Kaluza--Klein (KK) excitations.
This construction must be carefully distinguished from the genuine \emph{compactification} of the 6d $\cN=(2,0)$ theory on the 4-manifold
$C_{g_1,n} \times C_{g_2}$.
As emphasized in~\cite{Gukov:2020btk,Gukov:2025dol}, the resulting 2d theory
$\cT_{G}[C_{g_1,n} \times C_{g_2}]$ depends crucially on the full KK spectrum
and the choice of polarization of the 4-manifold.
In particular, its partition function requires summing over the allowed magnetic
flux sectors on $C_{g_1,n}$ and $C_{g_2}$~\cite{Benini:2016hjo,Closset:2017bse}. Moreover, there may be additional SCFT fixed points on the entire moduli space. 
Therefore, the theory studied in this work is \emph{not} identical to the compactified theory $\cT_{G}[C_{g_1,n} \times C_{g_2}]$.

An immediate consequence is that our 2d theory does
\emph{not} exhibit an exchange symmetry between $C_{g_1}$ and $C_{g_2}$ when there is no puncture.
Such a symmetry may exist only in the full compactification with
KK modes and flux sectors included; in the zero-mode truncation, this symmetry is not expected to be present, and our results indeed confirm its absence.

\bigskip

Central charges of a 2d $\cN=(0,4)$ theory are determined by its  't~Hooft anomalies. In particular, the right-moving central charge is fixed by the ’t~Hooft anomaly of the $U(1)$ R-symmetry of the $\cN=(0,2)$ subalgebra and is given by
\begin{equation}
\label{c_R}
c_R = 3 k_R = 3\Tr \gamma_3 R^2~.
\end{equation}
Here, $\gamma_3$ is the 2d chirality matrix, and the trace is evaluated over all Weyl fermions in the theory.
The difference between the left- and right-moving central charges is determined by the gravitational anomaly,
\begin{equation}
\label{grav-anomaly}
k_g \equiv c_L - c_R = \Tr \gamma_3 ~. 
\end{equation}

When the superconformal R-symmetry of the infrared (IR) fixed point is directly inherited from the ultraviolet (UV) theory, the central charges can be computed straightforwardly from UV anomaly data using \eqref{c_R} and \eqref{grav-anomaly}. However, an emergent R-symmetry may arise in the infrared that is not visible in the ultraviolet description. When such an emergent symmetry is present, a naive computation based solely on UV data is no longer valid.
More generally, the values of the central charges are sensitive to IR dynamics, in particular to the structure of the vacuum moduli space and to the fate of gauge degrees of freedom along the renormalization group flow. A careful analysis of these IR effects is therefore essential. The main objective of this paper is to clarify how the correct IR R-symmetry is identified, and how the corresponding central charges should be computed.

In the IR, the theory flows to an NLSM whose target space is a specific branch of the vacuum moduli space. The resulting sigma model realizes a 2d small $\cN=(0,4)$ superconformal algebra, equipped with an affine $SU(2)_{\mathrm{IR}}$ R-symmetry at level $k = c_R/6$. This level is equal to the quaternionic dimension of the sigma-model target space.
Since different branches of the vacuum moduli space may preserve different R-symmetries, the value of $c_R$ generally depends on the choice of IR R-symmetry. A key constraint \cite{Witten:1997yu} is that the IR R-symmetry must act trivially on the scalar fields parametrizing the moduli space. Consequently, the correct IR R-symmetry is one under which all scalar fields are neutral, while the fermions transform with charges compatible with $\cN=(0,4)$ supersymmetry. From the UV perspective, there are two natural candidates for such an IR R-symmetry.

\paragraph{Choice 1.} Let us first take the IR R-symmetry to be 
 $SU(2)_{+}$. With this choice, the left-moving fermions 
\((\lambda_-,\tilde{\lambda}_-)\) in the $(0,4)$ vector multiplet and the right-moving fermions 
\((\psi_+,\tilde{\psi}_+)\) in the $(0,4)$ hypermultiplet carry non-trivial IR R-charge, according to Table \ref{tab:UV-charges}.  
Their contributions give
\begin{equation}\label{eq:S2-naive-central-charge}
    c_R^{(0)}= 3\Tr \gamma_3 R^2_+= 3\times 2 \times (n_h-n_v)=6(n_h-n_v)\ .
\end{equation}
However, this expression is correct only for the sphere reduction 
$g_1=g_2=0$, with an arbitrary number of punctures $n$, and it describes the right-moving central charge of the NLSM on the Higgs branch that is parametrized by hypermultiplet scalars.  
For more general values of $(g_1,g_2)$, this result fails to reproduce the actual IR central charge due to unbroken gauge groups, and further modifications are required, as will be explained in Section \ref{subsec:conjecture}.

\paragraph{Choice 2.} Alternatively, we can take
 $SU(2)_{-}$ as the IR $\cN=(0,4)$ R-symmetry. In this case, the left-moving fermions $(\lambda_-,\tilde{\lambda}_-)$ in the $(0,4)$ vector multiplet and the right-moving fermions $(\lambda_+, \tilde{\lambda}_+)$ in the $(0,4)$ twisted hypermultiplet have non-trivial IR R-charges. The resulting right-moving central charge is
\begin{equation}\label{cR-tw}
        c^{\text{tw}}_R = 3\Tr \gamma_3 R^2_-= 3\times 2\times (g_2 n_v -n_v)=6(g_2-1)n_v\ ,
    \end{equation}
For $g_2 \ge 2$, this precisely matches the central charge of the NLSM on the twisted Higgs branch, where the branch is parametrized by twisted hypermultiplet scalars, as we will see in \S\ref{subsubsec:twisted-Higgs}.
While this branch exists for $g_2 = 1$ since the twisted hypermultiplets exist in the theory, additional subtleties arise, and this formula must be modified accordingly.

\bigskip

The gravitational anomaly is given by the difference between the numbers of the left-moving and right-moving fermions:
\begin{equation}\label{gravitational-anomaly}
    k_g =\Tr \gamma_3= 2(n_v + g_2 n_h)-2(n_h+g_2n_v)= 2(g_2-1)(n_h-n_v)\ .
\end{equation}
Note that for $g_2=1$, the resulting 2d theory has $\cN=(4,4)$ supersymmetry and the gravitational anomaly vanishes as expected.

\paragraph{Issue on the central charges.} Consider first the simplest situation in which the first Riemann surface has no punctures, i.e., $n=0$. Then, the naive attempt to compute the right-moving central charge as the 't Hooft anomaly $k_{SU(2)_+}$ 
leads to the expression \eqref{eq:S2-naive-central-charge}, which is
\begin{equation}
c_R^{(0)}
=
6(1-g_1)r_G~.
\end{equation}
However, this expression becomes negative when $g_1$ is greater than one, which is incompatible with the unitarity of the resulting 2d theory. Furthermore, when $g_1=1$, $c_R^{(0)}$ vanishes although the theory is non-trivial. 
This already indicates that the expression \eqref{eq:S2-naive-central-charge} is incorrect for a general $g_1$. 

The naive evaluation of the ’t~Hooft anomaly $k_{SU(2)_+}$ fails to reproduce the physically correct right-moving central charge for $g_1>0$. The underlying reason is that, at a generic point on the Higgs branch, the Cartan subgroup of the gauge group remains unbroken~\cite{Hanany:2010qu}.  
In two dimensions, however, an Abelian vector multiplet is \emph{gapped} in the infrared~\cite{Witten:1997yu,Delmastro:2021otj}. Consequently, the Cartan part of gauginos does \emph{not} contribute to the right-moving central charge~\cite{Putrov:2015jpa,Nawata:2023aoq,Jiang:2024ifv}.  
More generally, when a subgroup of the gauge group remains unbroken at a generic point on a branch of the vacuum moduli space, the superconformal $SU(2)$ R-symmetry in the IR theory is \emph{emergent} and is not manifest in the UV theory. As a result, performing ’t~Hooft anomaly matching using a UV R-symmetry and the full UV field content does \emph{not} yield the correct central charges.
\footnote{Such situations commonly arise when the 6d $(2,0)$ theory is compactified on a four-manifold $M_4$ with non-trivial one-cycles, i.e. the first Betti number $b_1(M_4)\neq 0$, leading to residual Abelian gauge sectors in the effective 2d theory.}  

In addition, as we will see below, the scalar fields in the hypermultiplet and twisted hypermultiplet sectors generally mix and acquire vacuum expectation values on the (special) Higgs branch. This mixing must be properly accounted for to determine the physically correct central charges. 

The same issue appears in the naive evaluation of the ’t~Hooft anomaly $k_{SU(2)_-}$. Specifically, the expression~\eqref{cR-tw} vanishes for $g_2=1$, even though the reduction of a class~$\cS$ theory on a torus yields a non-trivial 2d theory. As we will explain shortly, this apparent discrepancy is again attributed to the presence of an unbroken gauge group.  

We will present a conjecture for the correct right-moving central charge and provide a detailed explanation of the discrepancy between the naive and physical results in Section~\ref{subsec:conjecture}.

\subsection{Anomaly polynomial analysis and its limitation}\label{sec:anomaly-poly}

Central charges can be read from the anomaly coefficients encoded in the anomaly polynomial. For a 2d theory with at least $\cN=(0,2)$ supersymmetry, the anomaly polynomial~\cite{Benini:2013cda} is given by
\begin{equation}
    A_4 = \frac{k_g}{24} p_1(T\cM_2)+\frac{c_R}{6}c_1(r_{2d})^2~,
\end{equation}
where $p_1(T\cM_2)$ is the first Pontryagin class of the tangent bundle to the 2d manifold $\cM_2$, $c_1(r_{2d})$ is the first Chern class for the $U(1)_{r_{2d}}$ R-symmetry bundle of $(0,2)$ superalgebra.

\paragraph{From 2d field contents.} One can construct the 2d anomaly polynomial by computing the relevant 't Hooft anomaly coefficients directly from the UV field content and the charges listed in Table~\ref{tab:(0,4)-fields}. 
The ’t~Hooft anomaly $k_F$ associated with a global symmetry $F$ is computed through the relation
\begin{equation}
\operatorname{Tr}\!\left(\gamma_3\, f^a f^b\right)
\;=\;
k_F\, \delta^{ab} \, .
\label{eq:kF_def}
\end{equation}
where $f^a$ denote the generators of the Lie algebra $\mathfrak{f}$ of $F$, and the trace is taken over all Weyl fermions in the theory. Each fermion contributes with a sign determined by its chirality and a weight determined by its representation under $F$.

Applying this prescription, one finds the (UV) anomaly polynomial~\footnote{For simplicity, we suppress the anomalies from flavor symmetries and other possible mixed anomalies.}
\begin{equation}\label{eq:2d-ap}
     A_4=(g_2-1)\left[
    (n_h-n_v)\left(
    \frac{1}{12}p_1(T\cM_2)-n_v c_2(R_-)+c_1(r)^2
    \right)
    \right]- (n_h -n_v) c_2(R_+)
\end{equation}
where $r$ denotes the background gauge field strength of $U(1)_r$ dimensionally reduced to 2d. 
Since the right-moving central charge is calculated in the same way as in \eqref{c_R}, the issue discussed above remains present in this approach.

\paragraph{From dimensional reduction.} 
Starting from the anomaly polynomial $A_8$ of the 6d $(2,0)$ SCFT, one obtains the anomaly polynomial $A_6$ of class~$\cS$ theories by appropriately integrating $A_8$ over the punctured Riemann surface $C_{g_1,n}$.  
For a resulting 4d $\cN=2$ theory with effective numbers of vector and hypermultiplets $(n_v,n_h)$, the anomaly polynomial takes the  form~\cite{Tachikawa:2015bga}
\begin{equation}\label{eq:4d-anomaly-polynomial} 
A_6 = (n_v-n_h)\left(
\frac{c_1(r)^3}{3}-\frac{c_1(r)}{12}p_1(T\cM_4)
\right)-n_v c_1(r)c_2(R) \ , 
\end{equation}
where $R$ and $r$ denote the $SU(2)_R$ and $U(1)_r$ R-symmetry bundles of the 4d $\cN=2$ theory.

To obtain the anomaly polynomial of the 2d theory that arises upon compactification on a second Riemann surface $C_{g_2}$, one further integrates $A_6$ over $C_{g_2}$. 
Preserving $\cN=(0,4)$ supersymmetry requires performing a topological twist along  $U(1)^{\text{tw}}_{C_{g_2}}=\mathrm{diag}\bigl( U(1)_{C_{g_2}} \times U(1)_r \bigr)$ as in \eqref{eq:U1r_twist}.  Under this twist, the 4d characteristic classes decompose into their 2d counterparts as
\begin{equation*}
    p_1(T\cM_4) = p_1(T\cM_2), \quad c_1(r)= c_1(F)+\frac{t}{2}
   ,\quad c_2(R)=-c_1(R_-)^2  \ ,
\end{equation*}
where $t$ is the Chern root of the tangent bundle of $C_{g_2}$, normalized such that  
 $\int_{C_{g_2}} t=2(1-g_2)$.
Integrating~\eqref{eq:4d-anomaly-polynomial} on $C_{g_2}$ gives the 2d $\cN=(0,4)$ anomaly polynomial,
\begin{equation}\label{eq:2d-anomaly-polynomial-from-4d}
   A_4= (g_2-1)\left[ (n_h-n_v)\left(c_1(F)^2-\frac{p_1(T\cM_2)}{12}\right)-n_v  c_1(R_-)^2\right]\ .
\end{equation}
The dimensional-reduction result~\eqref{eq:2d-anomaly-polynomial-from-4d} reproduces~\eqref{eq:2d-ap} except for the anomaly term of $SU(2)_{+}$. 
While the Cartan subgroup of $SU(2)_+$ is identified with $\mathrm{diag} \left(U(1)_{C_{g_2}}  \times U(1)^{-1}_r \right)$ \eqref{eq:2d-R-identification}, the characteristic class $t$ of $C_{g_2}$ is completely integrated out during the dimensional reduction. 
As a result, the $SU(2)_+$ anomaly cannot be detected by integrating the 4d anomaly polynomial, leading to its absence in~\eqref{eq:2d-anomaly-polynomial-from-4d}.

The general lesson from this analysis is the following.  
When a 6d $(2,0)$ SCFT is compactified on a 4-manifold with appropriate topological twists, the standard procedure of integrating its anomaly polynomial $A_8$ over the 4-manifold does \emph{not} necessarily yield the correct central charge of the resulting 2d theory.  
Depending on the choice of topological twist and, crucially, on how the superconformal R-symmetry is realized in the infrared, part of the information relevant to the IR central charge may be integrated out along the way.

\subsection{Conjectures on central charges}\label{subsec:conjecture}

With the limitations of both the anomaly polynomial approach and naive 't~Hooft matching established, we now turn to a direct infrared analysis. A 2d $(0,4)$ theory flows in the infrared to an NLSM on an irreducible component of the vacuum moduli space. In general, the vacuum moduli space has multiple irreducible components, and the affine $SU(2)$ superconformal R-symmetry depends on the choice of irreducible component. The right-moving central charge is determined by the quaternionic dimension of the target space of the non-linear sigma model. To count the number of remaining massless hypermultiplets and twisted hypermultiplets after Higgsing, it is useful to re-examine the spectrum from the viewpoint of the Higgs mechanism.
For gauge group $SU(2)$, a detailed analysis of the Higgsing mechanism in 2d (0,4) theories is given in Appendix~\ref{app:Higgs}, which supplements the discussion in this section.

\paragraph{Higgs Branch.}

Let us first focus on the simplest case with $g_{1}=0$.  
Turning on generic VEVs for the hypermultiplet scalars, $\langle q\rangle$ and $\langle \tilde q\rangle$, completely breaks the gauge group~\cite{Hanany:2010qu}. 
As a result, all $n_v$ gauge fields acquire masses, and supersymmetry then requires the gauginos, which are left-moving Weyl fermions, to become massive as well.
However, in two dimensions, a single left-moving Weyl fermion cannot acquire a mass on its own: a mass term necessarily couples a left-moving fermion to a right-moving one. Indeed, a generic fermion mass term in two dimensions has the form of
\begin{equation} 
S =\int d^{2}x\, \left(\ii\psi_{+}^{\dagger}\partial_{-}\psi_{+} + \ii\psi_{-}^{\dagger}\partial_{+}\psi_{-} \right)- m\left(\psi_{-}^{\dagger}\psi_{+} + \psi_{+}^{\dagger}\psi_{-}\right)
\end{equation}
which explicitly shows that a massive Dirac fermion requires both chiralities. 
On the Higgs branch, the right-moving partners come from the hypermultiplets. Thus, the $n_v$ left-moving gauginos pair up with $n_v$ right-moving fermions from the hypermultiplets, rendering $n_v$ hypermultiplets massive.  
After integrating out these massive fields, the low-energy theory contains $(n_h - n_v)$ massless hypermultiplets, which reproduces the naive expression \eqref{eq:S2-naive-central-charge} for the right-moving central charge. Since the gauge group is completely broken on this branch, no residual Abelian sector survives in the infrared, and formula \eqref{eq:S2-naive-central-charge} therefore yields the correct right-moving central charge.

We now turn to the cases with $g_{1}>0$.  
As emphasized in~\cite{Hanany:2010qu}, when $g_{1}>0$ the Higgs branch structure changes qualitatively: 
at a generic point on the Higgs branch, the Cartan subgroup of the gauge group of the parent 4d theory remains unbroken. 
As a result, an Abelian gauge sector
$U(1)^{\,r_G g_1}$
survives in the 2d theory.
Consequently, only
$n_v - g_{1} r_{G}$
vector multiplets become massive via the Higgs mechanism.
By the same reasoning as in the $g_{1}=0$ case, the fermionic partners of these massive vector multiplets must pair with right-moving fermions from hypermultiplets, as shown in Appendix \ref{app:Higgs}. 
This renders $(n_v - g_{1} r_{G})$ hypermultiplets massive, leaving
$n_h - (n_v - g_{1} r_{G})$
massless hypermultiplets in the low-energy spectrum.

In addition, the presence of the unbroken gauge group $U(1)^{g_{1} r_{G}}$ has an important further consequence. 
Upon reduction on the second Riemann surface $C_{g_{2}}$, each unbroken $U(1)$ factor gives rise to  $g_2$ massless twisted hypermultiplets in two dimensions.
Altogether, this leads to
$g_{1} g_{2} r_{G}$
massless twisted hypermultiplets in the effective 2d theory.

In two dimensions, the residual Abelian vector multiplets are \emph{gapped} in the deep infrared~\cite{Witten:1997yu,Delmastro:2021otj}, and gapped degrees of freedom do \emph{not} contribute to ’t Hooft anomalies of the IR fixed point. Consequently, combining the contributions from massless hypermultiplets and twisted hypermultiplets, the deep IR spectrum contains
$n_h - (n_v - g_{1} r_{G}) + g_{1} g_{2}r_{G}$
massless bosonic degrees of freedom. This branch is characterized by the maximum number of massless hypermultiplet scalars, and we call this branch the \emph{special Higgs branch}. A summary of the massless field content on this branch is given in
Table~\ref{tab:fields_on_special}.
Based on this spectrum, we propose that the right-moving central charge of the
special Higgs branch CFT is
\begin{equation}\label{eq:conjecture}
\begin{aligned}
   c_R &= c_R^{(0)} + 6 g_1 (1+g_2) r_G=6\left(n_h-n_v+g_1(1+g_2)r_G\right)~. \\ 
\end{aligned}
\end{equation}

In particular, for the case $G=SU(2)$ and $r_G=1$, one has $n_h-n_v=1-g_1+n$, so the conjecture becomes
\begin{equation}\label{eq:cc-of-special-Higgs}
    c_R=6(g_1g_2+n+1)~,
\end{equation}
which will be confirmed by Hilbert series computations in Section \ref{subsubsec:special-Higgs}.
As additional supporting evidence, we study the dimensional reduction of a single M5 brane on $C_{g_1,n} \times C_{g_2}$ in Appendix \ref{app:single_M5}. Since the  worldvolume theory is Abelian, the 2d $\cN=(0,4)$ theory after reduction is free. Its central charge can be determined by counting the
positive contributions from the bosonic and fermionic degrees of freedom, providing an independent consistency check of \eqref{eq:cc-of-special-Higgs}.

\begin{table}[ht]
\renewcommand{\arraystretch}{1.3}
    \centering
    \begin{tabular}{c|c|c|c}
    \hline
      fields & UV  & $\langle q \rangle , \langle \tilde{q} \rangle$ &  IR  \\
      \hline
        $U=(A_\mu,\lambda_-),~\Theta=(\tilde\lambda_-)$ & $n_v$ & $g_1 r_G$ & $g_1 r_G ~(\text{gapped})$ \\
        $\Sigma=(\sigma, \lambda_+^\dagger),~\tilde\Sigma=(\tilde\sigma,\tilde\lambda_+)$ & $g_2 n_v$ & $g_1 g_2 r_G$ & $g_1 g_2 r_G$ \\
        $\Phi=(q,\psi_+),~\tilde\Phi=(\tilde q,\tilde \psi_+)$ & $n_h$ & $n_h-n_v+g_1 r_G$ & $n_h-n_v+g_1 r_G$ \\
        $\Gamma=(\psi_-),~\tilde\Gamma=(\tilde\psi_-)$ & $g_2 n_h$ & $g_2(n_h-n_v+g_1 r_{G})$ & $g_2(n_h-n_v+g_1 r_{G})$ \\
        \hline
    \end{tabular}
    \caption{Number of massless effective fields of 2d $\cN=(0,4)$ rank-1 gauge theory in special Higgs branch at three energy scales.}
    \label{tab:fields_on_special}
\end{table}

\paragraph{Twisted Higgs Branch for $g_2=1$ and $G=SU(2)$.}
We now turn to the analysis of the twisted Higgs branch, starting with the
simplest case in which $g_2=1$ and the gauge group is $G=SU(2)$.
Upon dimensional reduction on a two-torus, a 4d
$\mathcal{N}=2$ gauge theory reduces to a 2d
$\mathcal{N}=(4,4)$ gauge theory with the same matter content.
More precisely, a 4d $\mathcal{N}=2$ vector multiplet (resp.
hypermultiplet) descends to a 2d $\mathcal{N}=(4,4)$ vector multiplet
(resp. hypermultiplet).
Recall that a 2d $\mathcal{N}=(4,4)$ vector multiplet can be
decomposed into an $\mathcal{N}=(0,4)$ vector multiplet $U$ together with an
$\mathcal{N}=(0,4)$ twisted hypermultiplet $\Sigma$.
Then, turning on VEVs for the scalar fields
$\langle \sigma \rangle$ and $\langle \tilde{\sigma} \rangle$ in $\Sigma$
parametrizes the $\mathcal{N}=(4,4)$ Coulomb branch.

As in 4d $\mathcal{N}=2$ gauge theories, the Coulomb branch is
characterized by the breaking of the non-Abelian gauge symmetry down to its
Cartan subgroup. In the present case, this results in an unbroken
$U(1)^{n_v/3}$ gauge symmetry, where each $U(1)$ factor corresponds to the
Cartan of a gauge node. All (0,4) hypermultiplets become massive on this branch. Table \ref{tab:fields_on_twisted} summarizes the effective degrees of freedom. 
As a result, the (0,4) twisted Higgs branch (the (4,4) Coulomb branch) is parametrized by scalars from $n_v/3$ massless (0,4) twisted hypermultiplets, leading to the right-moving central charge 
\begin{equation}\label{tw-g2=1}
    c_R = 6\times \frac{n_v}{3} = 2n_v~.
\end{equation}
This will be verified in Section \ref{subsubsec:twisted-Higgs}.

For higher-rank theories, the dimension of the twisted Higgs branch is
expected to depend on the detailed data of the punctures.
However, in the absence of a Lagrangian description for generic higher-rank
class $\mathcal{S}$ theories, we are currently unable to formulate a precise
conjecture for the dimension of the twisted Higgs branch when $g_2=1$.
We therefore leave a systematic analysis of this problem to future work.

\begin{table}[ht]
\renewcommand{\arraystretch}{1.3}
    \centering
    \begin{tabular}{c|c|c|c}
    \hline
      fields & UV  & $\langle \sigma \rangle ,\langle \tilde{\sigma} \rangle$ &  IR  \\
      \hline
        $U=(A_\mu,\lambda_-),~\Theta=(\tilde\lambda_-)$ & $n_v$ & $n_v/3$ & $n_v/3 ~(\text{gapped})$ \\
        $\Sigma=(\sigma, \lambda_+^\dagger),~\tilde\Sigma=(\tilde\sigma,\tilde\lambda_+)$ & $ n_v$ & $n_v/3$ & $n_v/3$ \\
        $\Phi=(q,\psi_+),~\tilde\Phi=(\tilde q,\tilde \psi_+)$ & $n_h$ & $0$ & $0$ \\
        $\Gamma=(\psi_-),~\tilde\Gamma=(\tilde\psi_-)$ & $ n_h$ & $0$ & $0$ \\
        \hline
    \end{tabular}
    \caption{Number of massless effective fields of 2d $\cN=(4,4)$ gauge theory with $g_2=1, G=SU(2)$ in twisted Higgs branch, analogous to the Coulomb branch structure of the corresponding 4d class $\mathcal{S}$ theory.}
    \label{tab:fields_on_twisted}
\end{table}

\paragraph{Twisted Higgs Branch for $g_2 \geq 2$.}

For $g_2 \geq 2$, the twisted Higgs branch preserves
$\mathcal{N}=(0,4)$ supersymmetry.
As will be discussed in Section~\ref{subsubsec:twisted-Higgs}, all gauge symmetries are
completely broken on this branch by the Higgs mechanism, triggered by VEVs of the twisted hypermultiplet scalars.
Consequently, all gauginos acquire
a mass by pairing with the right-moving fermions in the $n_v$ twisted hypermultiplets.
As in the $\mathcal{N}=(4,4)$ case, all standard (0,4) hypermultiplets are also integrated out (or lifted) on the twisted Higgs branch.
Therefore, after Higgsing, a simple counting shows that the only remaining massless degrees of freedom are
$(g_2-1) n_v$ massless twisted hypermultiplets.
These fields parametrize the twisted Higgs branch, which is the target space of the NLSM in the infrared. 

The contribution of these massless twisted hypermultiplets precisely reproduces
the right-moving central charge \eqref{cR-tw} obtained from the ’t~Hooft anomaly of the
$SU(2)_-$ R-symmetry.
This agreement provides a direct consistency check of the right-moving central charge \eqref{cR-tw} via the Higgs mechanism.


\section{Vacuum moduli spaces for \texorpdfstring{$G=SU(2)$}{G=SU(2)}}\label{sec:vac-moduli}

As a consistency check of the conjectures in \eqref{eq:conjecture} and \eqref{tw-g2=1}, we now present a detailed analysis of the infrared vacuum structure for theories with gauge group $G=SU(2)$, for which an explicit ultraviolet Lagrangian description is available.

By solving the vacuum equations explicitly, we analyze the structure of the vacuum moduli space, which generically decomposes into several irreducible components. Among these, we systematically study two distinguished components—the special Higgs branch and the twisted Higgs branch—by computing their Hilbert series, which encode the counting of holomorphic functions on the corresponding branches. The Hilbert series of these two components exhibit different behaviors as the parameters $g_1$ and $n$ are varied. In particular, we develop a gluing method that allows us to compute the Hilbert series of the special Higgs branch for general values of $(g_1,n)$. By contrast, the Hilbert series of the twisted Higgs branch typically factorizes into contributions associated with individual gauge nodes.

From these Hilbert series, we extract the dimensions of the moduli spaces and the central charges of the IR NLSM on each branch, finding excellent agreement with the conjectured values. We also determine the pattern of unbroken gauge symmetries and propose an interpretation of the IR R-symmetries. Specifically, we conjecture that the superconformal R-symmetry in the infrared arises from a nontrivial mixing between a UV R-symmetry and an emergent global symmetry that becomes manifest only at low energies.

\subsection{(0,4) Lagrangian}\label{subsec:field-content}

We now derive the effective $(0,4)$ Lagrangian from the dimensional reduction of the 4d theory $\mathcal{T}_{SU(2)}[C_{g_1,n}]$ on $C_{g_2}$. For notational simplicity, we omit the gauge group label and write the theory simply as $\mathcal{T}[C_{g_{1},n}]$. We shall sometimes denote the numbers of gauge nodes and trinions by
\begin{equation}
    N_v=\frac{n_v}{3}=3(g_1-1)+n~,\quad N_h=\frac{n_h}{4}=2(g_1-1)+n~.
\end{equation}

Upon dimensional reduction, the 4d superpotential specifically yields $(0,2)$-type couplings with $J$- and $E$-terms in two dimensions.
Solving the resulting $J$-, $E$-equations subject to $D$-term constraints defines a hyper-Kähler vacuum moduli space, which we refer to as the \emph{full Higgs branch} of the 2d theory. 
In general, this vacuum moduli space decomposes into multiple irreducible components.  
Our goal in what follows is to analyze the structure of this moduli space.

To set the stage, recall the cubic superpotential of the 4d theory $\mathcal{T}[C_{g_1,n}]$,
\begin{equation}\label{4dsuperpotential}
    \mathcal L_{W_{\mathrm{4d}}}
    = \tilde{Q}_{\mathrm{4d}}\,\Phi_{\mathrm{4d}}\,Q_{\mathrm{4d}}
    + \mathrm{h.c.}~.
\end{equation}
Upon reduction on $C_{g_2}$, this superpotential yields an effective 2d interaction written naturally in terms of $(0,2)$ multiplets~\cite{Tong:2014yna,Putrov:2015jpa}:
\begin{equation}
    \begin{aligned}
    \mathcal L_{W_{\mathrm{2d}}}&=\tilde{Q} \Theta Q + 2\ii\sum_{j=1}^{g_2}\left(\tilde{\Gamma}_j \tilde{\Sigma}_j Q +  \tilde{Q} \tilde{\Sigma}_j\Gamma_j\right)+\mathrm{h.c.}~\\
    &=Q_{abc}Q_{\ap\bp\cp}\Theta^{a\ap}\epsilon^{b\bp}\epsilon^{c\cp}-\ii \sum_{j=1}^{g_2}\Gamma_{jabc}Q_{a^\prime b^\prime c^\prime}\tilde \Sigma_j^{aa^\prime} \epsilon^{bb^\prime}\epsilon^{cc^\prime}+\mathrm{h.c.}~,
    \end{aligned}    
\end{equation}
where $\epsilon=\begin{psmallmatrix}0 & 1 \\ -1 & 0\end{psmallmatrix}$, and our conventions for the $(0,2)$ multiplets are summarized in Appendix~\ref{app:notations}.

From the superpotential, we can deduce the holomorphic $J$-functions associated with each $(0,2)$ Fermi multiplet:
\begin{equation}
    {J}_{\Theta^{a\ap}}=Q_{abc}Q_{\ap\bp\cp}\epsilon^{b\bp}\epsilon^{c\cp}~,\quad {J}_{\Gamma_{jabc}}=-\ii Q_{a^\prime b^\prime c^\prime}\tilde \Sigma_j^{aa^\prime} \epsilon^{bb^\prime}\epsilon^{cc^\prime}~.
\end{equation}
The compatible holomorphic $E$-functions for the Fermi multiplets are given by
\begin{equation}
{E}_{\Theta}=\frac{1}{2}\sum_{j=1}^{g_2}\left[\tilde\Sigma_j,\Sigma_j\right]~,\quad {E}_{\Gamma_{jabc}}=\ii Q_{a^\prime b^\prime c^\prime} \Sigma_j^{aa^\prime} \epsilon^{bb^\prime}\epsilon^{cc^\prime}~,
\end{equation}
which are subject to the constraints:
\begin{equation}
    E\cdot J=E_\Theta J_{\Theta}+\sum_{j=1}^{g_2}E_{\Gamma_j}J_{\Gamma_j}=0~.
\end{equation}

The supersymmetric vacua are obtained by imposing the vanishing of all holomorphic $J$- and $E$-terms, together with the standard $D$-term constraints. Concretely, the scalar fields $q_{abc}$ in the hypermultiplets $Q_{abc}$ and the scalars $\sigma_j,\tilde{\sigma}_j$ in the twisted hypermultiplets $\Sigma_j,\tilde{\Sigma}_j$ for $j=1,\ldots,g_2$ must satisfy  
\begin{equation}\label{general_vac_eqn}
J_\Theta(q)
= J_{\Gamma_j}(q,\sigma,\tilde{\sigma})
= J_{\tilde{\Gamma}_j}(q,\sigma,\tilde{\sigma})
= E_\Theta(\sigma,\tilde{\sigma})
= E_{\Gamma_j}(q,\sigma,\tilde{\sigma})
= E_{\tilde{\Gamma}_j}(q,\sigma,\tilde{\sigma})
= 0~.
\end{equation}
After imposing the $D$-term conditions, the solution is expected to describe a union of hyper-Kähler cones, providing the full vacuum moduli space of the 2d $\cN=(0,4)$ theory that we analyze in the following subsections.

\subsection{Vacuum moduli spaces and Hilbert series}\label{subsec:hilbert-series}

The structure of the Higgs branches of class $\mathcal{S}$ theories of type $A_1$ has been analyzed in~\cite{Hanany:2010qu} by explicitly calculating the Hilbert series. 
We perform a similar analysis for the vacuum moduli spaces determined by the $J$-term and $E$-term equations~\eqref{general_vac_eqn} of the  2d $(0,4)$ theories. The resulting vacuum moduli space typically decomposes into several irreducible components, each with its own Hilbert series.\footnote{Both the identification of irreducible components and the computation of their Hilbert series can be performed using \texttt{Macaulay2}~\cite{M2}. See also~\cite{Cremonesi:2015lsa} for an illustrative example.} 

To compute the Hilbert series, we first construct the $F$-flat series
$F_{(g_1,n)}^\flat(t,z_a,x_j;g_2)$, a generating function encoding all
holomorphic operators on the $F$-flat space associated with the moduli space
of the theory. The variable $t$ is a fugacity tracking operator dimension
(or R-charge), $z_a$ $(a=1,\dots,N_v)$ are gauge fugacities associated with
the $N_v$ $SU(2)$ gauge nodes, and $x_j$ $(j=1,\dots,n)$ are flavor
fugacities for the $n$ punctures. The Hilbert series is then obtained by
projecting onto gauge-invariant operators via the Molien--Weyl
integral~\cite{Butti:2007jv,Hanany:2008kn,Gray:2008yu}:
\begin{equation}
    G_{(g_1,n)}(t,x_j;g_2)=\oint_{|z_a|=1}\prod_{a=1}^{N_v}
    d\mu_{SU(2)}(z_a)\, F_{(g_1,n)}^\flat(t,z_a,x_j;g_2)~,
\end{equation}
where $d\mu_{SU(2)}(z_a)$ denotes the Haar measure for the $a$-th $SU(2)$
gauge factor, explicitly given by
\begin{equation}
    \oint_{|z_a|=1} d\mu_{SU(2)}(z_a)=\frac{1}{2\pi \mathrm{i}}
    \oint_{|z_a|=1} dz_a \frac{1-z_a^2}{z_a}~.
\end{equation}
The contour integration implements the projection onto the trivial
representation of each gauge group, with the factor $(1-z_a^2)/z_a$ arising
from the $SU(2)$ Vandermonde determinant. The resulting Hilbert series
$G_{(g_1,n)}$ is therefore a function of the flavor fugacities $x_j$ alone,
graded by $t$. The degree in $t$ counts the charge under the UV R-symmetry
$\mathrm{diag}(SU(2)_+\times SU(2)_-)$, so that the coefficient of $t^k$
enumerates gauge-invariant holomorphic functions on the moduli space carrying
UV R-charge $k$.

The structure of irreducible components in the moduli space may vary across different duality frames.  
Nevertheless, there exist two distinguished components whose presence is \emph{frame-independent}.  
We refer to these as the \textit{special Higgs branch} and the \textit{twisted Higgs branch}, according to the types of scalar fields that parametrize them.\footnote{A systematic analysis of additional irreducible or embedded components is beyond the scope of this work and will be pursued in future investigations.}  
The special Higgs branch is parametrized by the scalar fields of hypermultiplets, and—when $g_{1}>0$— also by those of twisted hypermultiplets.  
In contrast, the twisted Higgs branch is generically parametrized entirely by the scalars arising from twisted hypermultiplets.  
Precise definitions of these branches will be introduced shortly, and explicit computational examples are provided below and in Appendix~\ref{app:hilbert-series}.

\paragraph{Comments on the elliptic genus.}
Another natural observable in 2d $\mathcal{N}=(0,4)$ theories is the
elliptic genus. As noted
in~\cite{Putrov:2015jpa,Nawata:2023aoq}, when $g_1=0$ and $g_2=0$, the elliptic genus
reduces to the Hilbert series of the (special) Higgs branch in the $q\to 0$
specialization. For $g_1>0$, however, this agreement breaks down for two
related reasons. First, the Hilbert series explicitly depends on the
superpotential --- the $F$-flat series is defined precisely by the
$F$-term relations --- whereas the elliptic genus is superpotential-independent.
Second, the two quantities are computed by qualitatively different contour
prescriptions: the Hilbert series employs the Molien--Weyl integral over the
unit circle, while the elliptic genus requires the Jeffrey--Kirwan residue
prescription. A further obstruction arises from the fact that, for $g_1>0$,
the Cardy limit of the elliptic genus does \emph{not} reproduce the correct
central charge of the theory. Consequently, the Hilbert series is the
appropriate observable for probing
the moduli space and verifying our conjecture for the central charge.

\subsubsection{Special Higgs branch}\label{subsubsec:special-Higgs}

\begin{figure}[ht]
    \centering
    \includegraphics[width=0.7\linewidth]{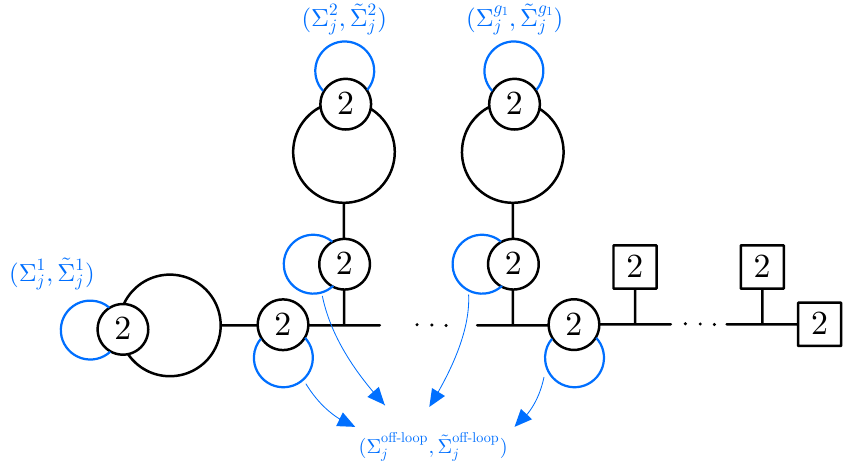}
    \caption{The tadpole frame of the case with generic $(g_1,n)$. The gauge nodes are of two types: one are on the loops of the quiver, and the others are not. The special Higgs branch requires the scalars of the twisted hypermultiplets $\left(\Sigma_j^{\text{off-loop}},\tilde\Sigma_j^{\text{off-loop}}\right)$ connecting to the nodes of the latter type must vanish.}
    \label{fig:std_frame}
\end{figure}

To define the special Higgs branch, it is convenient to work in a particular duality frame of $\cT[C_{g_1,n}]$ on $C_{g_2}$ obtained by gluing the basic building blocks—tadpoles ($C_{1,1}$ on $C_{g_{2}}$) and trinions ($C_{0,3}$ on $C_{g_{2}}$).  
We refer to this quiver presentation as the \textit{tadpole frame}.  
In this frame, each loop in the quiver diagram (each handle of the Riemann surface) has a single gauge node, as illustrated in Figure~\ref{fig:std_frame}.  
The gauge nodes naturally fall into two categories: the loop node from the tadpole, and the node from gluing two building blocks.  
For a surface of type $(g_{1},n)$, the tadpole frame therefore contains precisely $g_{1}$ loop gauge nodes.

\paragraph{Ideal structures.}

The ideal of the full Higgs branch, generated by the $J$-term and $E$-term equations~\eqref{general_vac_eqn}, can be decomposed into three classes of equations,
\begin{equation}\label{ideal_full_Higgs}
    f_{qq}=0 ~, \qquad
    f_{q\sigma}=0~, \qquad
    f_{\sigma\sigma}=0~.
\end{equation}
Here, $f_{qq}$ consists of equations involving only the hypermultiplet scalars $q$ and coincides with the familiar 4d $F$-term relations.  
The equations $f_{\sigma\sigma}$ involve solely the twisted hypermultiplet scalars $(\sigma,\tilde{\sigma})$ while $f_{q\sigma}$ contains mixed terms coupling $q$ and $(\sigma,\tilde{\sigma})$.  
Explicit examples of these equations can be found in \eqref{JE_eqn(0,4)}, \eqref{JE_eqn(1,1)}, \eqref{Fterm_f2}, and \eqref{JE(2,0)}.

In the tadpole frame, the ideal defining the \emph{special Higgs branch} corresponds to one of the prime components of~\eqref{ideal_full_Higgs}. This component imposes additional constraints on $f_{qq}$, $f_{q\sigma}$, and $f_{\sigma\sigma}$. Since writing down the complete set of constraints explicitly is rather involved, we will not attempt to do so here.
Instead, we focus on the structure of the constrained $f_{\sigma\sigma}$ equations that characterize the special Higgs branch. Interestingly, these constraints take a simple and universal form, determined solely by the types of gauge nodes to which the twisted hypermultiplets are attached in the quiver.

Twisted hypermultiplets in the tadpole frame fall into two categories depending on whether their associated gauge nodes lie on a loop or not.  
Accordingly, the scalar VEVs on the special Higgs branch are subject to the following conditions:
\begin{itemize}
    \item The scalars in the twisted hypermultiplets should have vanishing commutator if they are associated with gauge nodes that are \textit{on a loop},
    \begin{equation}\label{constr_special_higgs_1}
    \left[\sigma^I_i,\sigma_j^I\right]=\left[\sigma^I_i,\tilde\sigma_j^I\right]=\left[\tilde\sigma^I_i,\tilde\sigma_j^I\right]=0~,\quad\forall~i,j=1,\cdots,g_2,~I=1,\cdots,g_1~.
\end{equation}
    These are precisely the conditions ensuring that the twisted hypermultiplet scalars parametrize a Cartan subalgebra along the loop directions.
\item  The scalars in the twisted multiplets should vanish if they are associated with gauge nodes that are \textit{not on a loop},
\begin{equation}\label{constr_special_higgs_2}
    \sigma_j^{\text{off-loop}}=\tilde\sigma_j^{\text{off-loop}}=0~,\quad\forall~i,j=1,\cdots,g_2~.
\end{equation}
    These fields therefore do not contribute to the parametrization of the special Higgs branch.
\end{itemize}

Depending on the geometric data, multiple duality frames may arise, e.g., cases with $g_1=0$, $n\geq6$, or $g_1=1$, $n\geq2$. 
In frames other than the tadpole frame, the defining equations for the special Higgs branch need not take the form of \eqref{constr_special_higgs_1} and \eqref{constr_special_higgs_2}.  
Nevertheless, we conjecture that the resulting {geometry} of the special Higgs branch remains invariant across these different frames.  
This frame independence can be checked by the agreement of the unrefined Hilbert series computed in each frame. For cases with $g_1=0$, the frame independence is inherited from the parent 4d $\cN=2$ theory, as the 2d vacuum equations for the special Higgs branch are exactly the same as that of the 4d theory. For cases with $g_1=1$,
we verified the frame independence at least for the case of $(g_1,n,g_2)=(1,2,1)$ (see \eqref{Fterm_f2}). Other cases exceed our current available computational capacity, but we conjecture that the frame-independence holds for general cases and leave the verification for future work.

As observed in~\cite{Hanany:2010qu}, the solutions to the hypermultiplet equations $f_{qq}=0$ leave $U(1)^{g_{1}}$ gauge group unbroken, where each $U(1)$ corresponds to an Abelian subgroup of an $SU(2)$ gauge node lying on a loop.  
Similarly, the constraints \eqref{constr_special_higgs_1} imply that, on the special Higgs branch of the tadpole frame, the twisted hypermultiplet scalars $\sigma^{I}_{i}$ associated with loop nodes acquire VEVs valued in a Cartan subalgebra of the corresponding $\fraksu(2)$. Moreover, the $f_{q\sigma}$ constraints require that, for each gauge node on a loop, the VEVs of $q$ and $\sigma$ take a parallel direction. For example, see \eqref{sH_constrain_(1,1)} for the case $(g_1,n)=(1,1)$. Hence the unbroken $U(1)$ is exactly the Cartan subgroup of the $SU(2)$ on the loop, generated by the VEVs $\sigma^I_i,\tilde\sigma_i^I$. Therefore, at the generic point, there is an unbroken gauge symmetry $U(1)^{g_{1}}$. 
Combining these observations with frame independence, we conclude that there is an unbroken gauge group $U(1)^{g_1}$ at the special Higgs branch.\footnote{Notice that gauge enhancement can happen at special loci. For example, at the origin, all the scalar fields vanishes, and hence the full gauge group is unbroken.}

\begin{table}[ht]
\centering
    \renewcommand{\arraystretch}{1.3}
\begin{tabular}{c|c|c}
\hline
{$n$} & {dim} & {Hilbert series}  \\
\hline
3 & 4 & $\frac{1}{(1-t)^8}$ \\
\hline
4 & 5 & $\frac{ (1 + t^2)(1 + 17 t^2 + 48 t^4 + 17 t^6 + t^8)}{\left(1-t^2\right)^{10}}$ \\
\hline
5 & 6 & $\frac{1+9 t^{2}+26 t^{3}+41 t^{4}+106 t^{5}+195 t^{6}+234 t^{7}+306 t^{8}+372 t^9+\text{...pal...}+t^{18}}{\left(1-t^2\right)^6 \left(1-t^3\right)^6}$  \\
\hline
6 & 7 & $\frac{1+11 t^{2}+118 t^{4}+538 t^{6}+1900 t^{8}+4109 t^{10}+6901 t^{12}+7804 t^{14}+\text{...pal...}+t^{28}}{\left(1-t^2\right)^7 \left(1-t^4\right)^7}$   \\
\hline
7 & 8 & \resizebox{0.83\textwidth}{!}{$\frac{\begin{pmatrix}
    1+13 t^{2}+85 t^{4}+120 t^{5}+377 t^{6}+792 t^{7}+1289 t^{8}+2904 t^{9}+4844 t^{10}+7736 t^{11}+12391 t^{12}\\
    +16568 t^{13}+24151 t^{14}+32328 t^{15}+38331 t^{16}+46416 t^{17}+51171 t^{18}+55056 t^{19}+59094 t^{20}\\+\text{...pal...}+t^{40}
\end{pmatrix}}{\left(1-t^2\right)^8 \left(1-t^5\right)^8}$} \\
\hline
\end{tabular}
\caption{The quaternionic dimensions and unrefined Hilbert series of the special Higgs branches of the case $(g_1,n)=(0,n)$. The results are identical to the Higgs branches of the corresponding 4d theories.}
\label{tab:HS_(0,n)}
\end{table}

\paragraph{Special cases.}
When $g_{1}=0$, the quiver contains no loops.  
Consequently, all twisted hypermultiplet scalars must vanish on the special Higgs branch.  In this situation, the $J$- and $E$-term equations of the 2d theory all vanish except for the one that is identical to the $F$-term equations of the 4d hypermultiplets. Thus, the special Higgs branch is parametrized entirely by the scalars of the 2d hypermultiplets and is \emph{identical} to the Higgs branch of the parent 4d theory studied in~\cite{Hanany:2010qu}, with quaternionic dimension $1+n$. The corresponding unrefined Hilbert series for various values of $n$ are summarized in Table~\ref{tab:HS_(0,n)}.

\paragraph{IR R-symmetry.}

We conjecture that, for $g_1>0$, there is an emergent global
symmetry $SU(2)_{\mathrm{new}}$ on the special Higgs branch in the infrared,
and the infrared R-symmetry group $SU(2)_{\mathrm{IR}}$ on this branch arises as a diagonal
combination of the ultraviolet R-symmetry $SU(2)_+$ and the emergent $SU(2)_{\text{new}}$: 
\be 
SU(2)_{\text{IR}}=\mathrm{diag}\left(SU(2)_+\times SU(2)_{\text{new}}\right)~.
\ee 
The emergent symmetry $SU(2)_{\mathrm{new}}$ acts non-trivially only on the
Cartan components of fields associated with gauge nodes \emph{on the loops} of
the quiver diagram.  More precisely, it acts on the Cartan parts
$(U^{\text{on-loop}}_{C}, \Theta^{\text{on-loop}}_{C})$ of the vector multiplets
and on the Cartan components
$(\Sigma^{\text{on-loop}}_{j,C}, \tilde{\Sigma}^{\text{on-loop}}_{j,C})$ of the
twisted hypermultiplets while acting trivially on all other multiplets.
The charges of the fields under the Cartan subgroups
$U(1)_{\mathrm{new}} \subset SU(2)_{\mathrm{new}}$ and
$U(1)_{\mathrm{IR}} \subset SU(2)_{\mathrm{IR}}$, in the tadpole frame,
are summarized in Table~\ref{tab:emergent}.
We emphasize that $SU(2)_{\mathrm{new}}$ is \emph{not} a symmetry of the UV gauge
theory since the superpotential is not invariant under the corresponding
charge assignments; it emerges \emph{only} in the IR NLSM.

\begin{table}[t]
\centering
\normalsize 
\renewcommand{\arraystretch}{1.3} 
\begin{tabular}{l|c|cc|c}
\hline
Superfield & numbers & $U(1)_+$ & $U(1)_{\text{new}}$ & $U(1)_{\text{IR}}$\\
\hline
$U^{\text{on-loop}}_C = (A_{\mu}, \lambda_{-})$ & $g_1 r_G$ & $\left(0,-\frac{1}{2}\right)$ & $\frac{1}{2}$ & $(\frac{1}{2},0)$ \\
$U^{\text{others}} = (A_{\mu}, \lambda_{-})$ & $n_v-g_1 r_G$ & $\left(0,-\frac{1}{2}\right)$ & 0 & $\left(0,-\frac{1}{2}\right)$ \\
$\Theta^{\text{on-loop}}_C=(\tilde{\lambda}_{-})$ & $g_1 r_G$ &  $-\frac{1}{2}$ & $\frac{1}{2}$ & 0 \\
$\Theta^{\text{others}}=(\tilde{\lambda}_{-})$ & $n_v-g_1 r_G$ &  $-\frac{1}{2}$ & 0 & $-\frac{1}{2}$ \\
$\Sigma^{\text{on-loop}}_{j,C}=(\sigma,\lambda_{+}^{\dagger})$ & $g_1g_2 r_G$ & $\left(\frac{1}{2},0\right)$ & $-\frac{1}{2}$ & $\left(0,-\frac{1}{2}\right)$ \\
$\Sigma_j^{\text{others}}=(\sigma,\lambda_{+}^{\dagger})$ & $(n_v-g_1r_G)g_2$ & $\left(\frac{1}{2},0\right)$ & 0 & $\left(\frac{1}{2},0\right)$ \\
$\tilde{\Sigma}^{\text{on-loop}}_{j,C}=(\tilde{\sigma},\tilde{\lambda}_{+})$ & $g_1g_2r_G$ & $\left(-\frac{1}{2},0\right)$ & $\frac{1}{2}$ & $\left(0,\frac{1}{2}\right)$ \\
$\tilde{\Sigma}_j^{\text{others}}=(\tilde{\sigma},\tilde{\lambda}_{+})$ & $(n_v-g_1r_G)g_2$ & $\left(-\frac{1}{2},0\right)$ & 0 & $\left(-\frac{1}{2},0\right)$ \\
\hline
$Q=(q,\psi_{+})$ & $n_h$ & $\left(0,\frac{1}{2}\right)$ & 0 & $\left(0,\frac{1}{2}\right)$ \\
$\tilde{Q}=(\tilde{q}$,$\tilde{\psi}_{+})$ & $n_h$ & $\left(0,\frac{1}{2}\right)$ & 0 & $\left(0,\frac{1}{2}\right)$ \\
$\Gamma_j=(\psi_{-})$ & $n_hg_2$ & 0 & 0 & 0\\
$\tilde{\Gamma}_j=(\tilde{\psi}_{-})$ & $n_hg_2$ & 0 & 0 & 0\\
\bottomrule
\end{tabular}
\caption{We conjecture that there is an emergent $SU(2)_{\text{new}}$ symmetry in the IR, and the IR R-symmetry $SU(2)_{\text{IR}}=\textrm{diag}(SU(2)_+\times SU(2)_{\text{new}})$. The charges of fields under the Cartans $U(1)_+$, $U(1)_{\text{new}}$ and $U(1)_{\text{IR}}$ are listed. In particular, $U(1)_{\text{new}}$ has non-zero charges only on the Cartan part of the vector multiplets and twisted hypermultiplets on a loop of the quiver. Note that this charge assignment is only for the tadpole frame as in Figure \ref{fig:std_frame}.}
\label{tab:emergent}
\end{table}

Since $SU(2)_{\mathrm{IR}}$ is an emergent symmetry that appears only at the
infrared fixed point, a correct computation of its anomaly coefficient must be
carried out using the IR field content listed in
Table~\ref{tab:fields_on_special}.
Performing the anomaly calculation with this spectrum, we obtain
\begin{equation}
    c_R=3\Tr \gamma_3 R^2_{\text{IR}}=3 \left[2\times g_1g_2r_G+2\times \left(n_h-n_v+g_1r_G\right)\right]=6\left(n_h-n_v+g_1(1+g_2)r_G\right)~,
\end{equation}
which precisely matches the conjecture~\eqref{eq:conjecture}.

As emphasized in Section~\ref{subsec:2d-fields}, a naive application of
’t~Hooft anomaly matching from the UV theory fails for the computation of the
right-moving central charge.
Nevertheless, although $SU(2)_{\mathrm{IR}}$ is \emph{not} a symmetry of the UV
theory, one may still formally compute the ’t~Hooft anomaly coefficient of $SU(2)_{\text{IR}}$ by using the charge assignments summarized in Table~\ref{tab:emergent}.
This formal computation yields
\begin{equation}
    c_R=3\Tr \gamma_3 R^2_{\text{IR}}=3 \left[-2\times\left(n_v-g_1r_G\right)+2\times g_1g_2r_G+2\times n_h\right]=6\left(n_h-n_v+g_1(1+g_2)r_G\right)~,
\end{equation}
in agreement with the IR calculation above. In this sense, the result may be viewed as a spurious form of
’t~Hooft anomaly matching: although the full $SU(2)_{\mathrm{IR}}$ symmetry is
emergent and absent in the UV description, the corresponding anomaly
coefficient can nevertheless be recovered from the UV data and reproduces the
correct infrared central charge.

In addition, this IR R-symmetry is also consistent with the requirement that scalars parametrizing the moduli space should be neutral under the IR R-symmetry~\cite{Witten:1997yu}. 
From the ideal structure, we see that the special Higgs branch is generically parametrized by the hypermultiplet scalars $q$ and the Cartan part of the twisted hypermultiplet scalars $\sigma^{\text{on-loop}}_{j,C}$ associated to the gauge nodes \emph{on the loops}. As shown in Table~\ref{tab:emergent}, these fields are indeed neutral under
$SU(2)_{\mathrm{IR}}$.

\paragraph{Gluing method.}

For $g_1\geq 1$, vacuum equations \eqref{constr_special_higgs_1},
\eqref{constr_special_higgs_2} and \eqref{general_vac_eqn} are complicated,
which makes direct computation of the Hilbert series difficult.
Instead, we use the \textit{gluing method} to compute the Hilbert series of
a larger quiver from those of simpler ones.
This method has been applied to the class~$\cS$ theories~\cite{Benvenuti:2010pq,Hanany:2010qu},
and we adapt it here to the 2d~$(0,4)$ context.

\begin{figure}[ht!]
    \centering
    \includegraphics[width=0.9\linewidth]{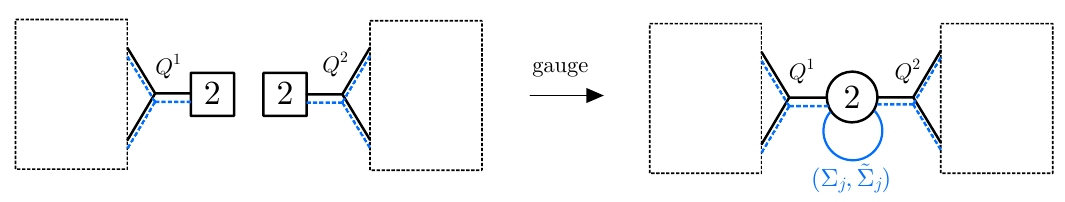}
    \caption{Gluing two quivers. The gluing will introduce 3 new $J$-term
    equations of $q^1,q^2$, and modify the $J$-term and $E$-term equations
    on both sides.}
    \label{fig:gluing}
\end{figure}

As is standard in the study of class~$\cS$ theories, a larger quiver can
be constructed by gluing two smaller, \textit{disjoint} quivers via gauging
a common flavor symmetry (see Figure~\ref{fig:gluing}).
The $J$-term and $E$-term equations of the resulting quiver are inherited
from those of the two original components, but the gluing procedure
introduces additional constraints that modify the vacuum equations.
More precisely, gluing leads to the following effects:
\begin{itemize}
    \item New twisted hypermultiplet scalars $\sigma_j$ and $\tilde{\sigma}_j$
    ($j=1,\ldots,g_2$) are introduced at the gauged node.

    \item Three new $J$-term equations involving the bifundamental fields
    $q^{1}$ and $q^{2}$ appear:
    \begin{equation}\label{gluing_eqn}
        q^{1}_{abc}\, q^{2}_{\ap\bp\cp}\,(e_{A})^{a\ap}\,
        \epsilon^{b\bp}\,\epsilon^{c\cp} = 0~,
        \qquad \forall~A = 1,2,3
    \end{equation}
    where $e_A$ is the canonical basis of adjoint representation,
    as defined by \eqref{canonical_basis}.

    \item The original $J$-term and $E$-term equations in each quiver
    acquire additional contributions of the form
    \be \label{modification}
        q^{I}_{\ap\bp\cp}\,\sigma_j^{a\ap}\,\epsilon^{b\bp}\epsilon^{c\cp}~,
        \qquad
        q^{I}_{\ap\bp\cp}\,\tilde{\sigma}_j^{a\ap}\,\epsilon^{b\bp}\epsilon^{c\cp}~.
    \ee
\end{itemize}

On the special Higgs branch, the fields $\sigma_j$ and $\tilde{\sigma}_j$
associated with gauge nodes not lying on a loop must vanish.
This follows directly from the structure of the vacuum equations: the
$J$- and $E$-term constraints force these scalars to zero whenever the
corresponding node is not part of a loop, so their vanishing is not an
additional assumption but a result of the equations themselves.
Consequently, for the special Higgs branch of the glued quiver, the
modifications~\eqref{modification} are absent, and the only non-trivial
new constraints introduced by the gluing procedure are the three
relations~\eqref{gluing_eqn}.
The vacuum equations therefore reduce exactly to those of the two original
quivers, supplemented solely by the additional gluing relations.

At the level of Hilbert series, this means that the $F$-flat generating
function of the special Higgs branch of the glued quiver is obtained by
\begin{equation}
    F_{H}^\flat(t,z)=G_{H,1}(t,z)\,G_{\text{glue}}(t,z)\,G_{H,2}(t,z)~,
\end{equation}
where $G_{H,1}(t,x_1)$ and $G_{H,2}(t,x_2)$ are the Hilbert series of the
special Higgs branches of the two original quivers, $x_1,x_2$ are the
flavor fugacities for the punctures being glued.
The gluing factor implementing the constraints~\eqref{gluing_eqn} is given by
\begin{equation}
    G_{\text{glue}}(t,z)=(1-t^2)(1-t^2z^2)(1-t^2z^{-2})~,
\end{equation}
where each factor corresponds to one of the three relations in
\eqref{gluing_eqn}, with $z$-weights determined by the $SU(2)$
representation content of the bifundamental fields.
This coincides with the gluing factor appearing in the 4d
analysis~\cite{Benvenuti:2010pq,Hanany:2010qu}.

The Hilbert series of the special Higgs branch is obtained by integrating
over the gauge fugacity:
\begin{equation}
    G_{H,(g_1+g'_1,\,n+n'-2)}(t;\,g_2)
    = \oint_{|z_{a}|=1} d\mu_{SU(2)}(z)\;
      G_{H,(g_1,n)}(t,z;\,g_2)\;
      G_{\mathrm{glue}}(t,z)\;
      G_{H,(g'_1,n')}(t,z;\,g_2)~,
\end{equation}
where $(g_1,n)$ and $(g_1^\prime, n^\prime)$ are the geometric data on the
two sides respectively.
Using this gluing method, one can compute the Hilbert series for any
configuration $(g_{1},n)$.
We have verified the results in two independent ways: by explicitly writing
and solving the full set of $J$- and $E$-term equations of the glued
quiver without invoking any factorization (see Appendix~\ref{app:hilbert-series}
for details), and by confirming frame independence, namely that different
quiver frames yield identical Hilbert series and hence identical
Higgs-branch geometries.

From the resulting Hilbert series, we deduce that the quaternionic
dimension of the special Higgs branch is
\begin{equation}\label{MH-dim}
    \dim_{\mathbb{H}} \mathcal{M}_{H}\!\left[C_{g_{1},n} \times C_{g_{2}}\right]
    = g_{1} g_{2} + n + 1~,
\end{equation}
leading to the right-moving central charge
\begin{equation}
    c_{R} = 6(g_{1}g_{2} + n + 1)~,
\end{equation}
which agrees with our conjecture~\eqref{eq:cc-of-special-Higgs}.

\paragraph{Non-palindromic Hilbert series for $g_1\geq 2$ and $g_{2}\geq 1$.}

From explicit computations of the Hilbert series, we find an intriguing phenomenon:  
for cases with $g_{1}\geq 2$ and $g_{2}\geq 1$, the numerators of the Hilbert series of the special Higgs branches are \emph{non-palindromic}.  
Concrete examples for $(g_{1},n)=(2,0)$ are presented in Table~\ref{tab:HS_(2,0)}, where we also verify that the corresponding ideal is minimal prime, and hence defines an irreducible and radical variety.

Such non-palindromic Hilbert series violate the Gorenstein property, i.e.,
\begin{equation}
    G(t^{-1}) = (-1)^{d}\, t^{a}\, G(t)~,
\end{equation}
for some integers $d$ and $a$.  
By Stanley's theorem~\cite{STANLEY197857}, this implies that the coordinate ring of the special Higgs branch is not Gorenstein (see also~\cite{Hanany:2016pfm}). This means that the special Higgs branch for $g_{1}\geq 2$ and $g_{2}\geq 1$ is no longer a symplectic singularity~\cite{beauville1999symplectic}.

The appearance of non-palindromic Hilbert series suggests that the geometry of the special Higgs branch undergoes a structural change for these values of $(g_{1},g_{2})$. 
At present, however, we do not have a clear geometric interpretation of these cases, and understanding the underlying structure remains an interesting open problem.

We further observe that for cases with $n=0$, the Hilbert series of the special Higgs branches are \emph{not} invariant under exchanging the two Riemann surfaces.  
Thus, the symmetry $g_{1}\leftrightarrow g_{2}$ is explicitly broken at the level of the Hilbert series, even though it is not apparent from the dimension formula~\eqref{MH-dim}.  
For example, using the gluing method, one finds
\begin{equation}
    G_{H,(2,0)}(t;3) \;\neq\; G_{H,(3,0)}(t;2)~,
\end{equation}
and notably, the numerators of both Hilbert series are non-palindromic.

\subsubsection{Twisted Higgs branch}\label{subsubsec:twisted-Higgs}

The twisted Higgs branch is parametrized purely by the twisted hypermultiplet scalar fields. 
It exists for $g_2 \geq 1$ and is characterized by the vacuum equations
\begin{equation}\label{eq_twisted_higgs}
q^I = 0~, 
\qquad 
\sum_{j=1}^{g_2}\bigl[\sigma^J_j,\tilde{\sigma}^J_j\bigr]=0~,
\qquad 
\forall~I=1,\dots,N_h~, \quad \forall~J=1,\dots,N_v~ .
\end{equation}
These equations define an ideal whose structure is inherently independent of the choice of duality frame.

For $g_2=1$, the vacuum equation reduces to
\begin{equation}
\bigl[\sigma^J,\tilde{\sigma}^J\bigr]=0
\end{equation}
at each gauge node. 
Consequently, the twisted hypermultiplet scalars can acquire vacuum expectation values valued in a Cartan subalgebra of $\fraksu(2)$. 
The resulting twisted Higgs branch is therefore given by
\begin{equation}\label{tw-torus}
\mathcal{M}_{\text{tw}}\!\left[C_{g_1,n}\times T^2\right]
\cong 
\bigl(\mathbb{H}/\mathbb{Z}_2\bigr)^{N_v} ,
\end{equation}
where the Weyl group $\mathbb{Z}_2$ of $SU(2)$ acts on the Cartan-valued VEVs. 
At a generic point on this branch, the Cartan subgroup of each gauge group remains unbroken, and the quaternionic dimension is consistent with the right-moving central charge \eqref{tw-g2=1}.

For $g_2>1$, although the gauge symmetry is completely broken at a generic point on the twisted Higgs branch, the vacuum equations \eqref{eq_twisted_higgs} at different gauge nodes are identical and decouple from one another. 
As a result, the twisted Higgs branch factorizes into a product of identical components,
\begin{equation}
\mathcal{M}_{\text{tw}}\!\left[C_{g_1,n}\times C_{g_2}\right]
\cong 
(\mathcal{M}_{\text{tw},g_2})^{\,N_v}~,
\end{equation}
where $\mathcal{M}_{\text{tw},g_2}$ denotes the twisted Higgs branch of quaternionic dimension $3(g_2-1)$ associated with a single gauge node.

We now compute the Hilbert series of the twisted Higgs branch.
Because of the reason described above, both the $F$-flat series and the Hilbert series factorize into independent contributions from each gauge node. From \eqref{eq_twisted_higgs}, the $F$-flat series takes the form
\begin{equation}
\label{eq:tw_Fflat_general}
    F^\flat_{\text{tw},(g_1,n)}(t,z_a;g_2)=\prod_{j=1}^{N_v}F^\flat_{\text{tw}}(t,z_a;g_2)~,
\end{equation}
where $z_a$ are the gauge fugacities, and $F^\flat_{\text{tw}}(t,z;g_2)$ denotes the $F$-flat series associated with a single gauge node.
Explicitly,
\begin{equation}
    F^\flat_{\text{tw}}(t,z;g_2)=\begin{dcases}
    \frac{2 t^3-t^2 \left(z^2+1+z^{-2}\right)+1}{(1-t)^2 \left(1-tz^{-2}\right)^2 \left(1-t z^2\right)^2}~, & g_2=1\\[3pt]
    \frac{\left(1-t^2\right) \left(1-{t^2}{z^{-2}}\right) \left(1-t^2 z^2\right)}{\left[(1-t) \left(1-{t}{z^{-2}}\right) \left(1-t z^2\right)\right]^{2g_2}}~, & g_2\geq 2~.
    \end{dcases}
\end{equation}
The Hilbert series is obtained by integrating over the gauge group at each node,
\begin{equation}\label{eq:tw_HS_general}
    G_{\text{tw},(g_1,n)}(t;g_2)=\prod_{j=1}^{N_v}\oint_{|z_a|=1}d\mu_{SU(2)}(z_a) F_{\text{tw}}^\flat(t,z_a;g_2)=\left(G_{\text{tw}}(t;g_2)\right)^{N_v}~.
\end{equation}
where $d\mu_{SU(2)}(z)$ denotes the Haar measure of $SU(2)$.
The single-node contribution $G_{\text{tw}}(t;g_2)$ is given by
\begin{equation}
    G_{\text{tw}}(t;g_2)=\begin{dcases}
        \frac{1+t^2}{(1-t^2)^2}~, & g_2=1\\
        \frac{1+\cdots}{(1-t^2)^{6g_2-6}(1+t^2)^{-1}}~, & g_2\geq 2~.
    \end{dcases}
\end{equation}
For the torus reduction $g_2=1$, the Hilbert series is that of $\bH/\bZ_2$, confirming the geometry \eqref{tw-torus} of the twisted Higgs branch.
Explicit expressions for $G_{\text{tw}}(t;g_2)$ with $g_2=1,\ldots,5$ are collected in Table~\ref{tab:HS_tw}.

From the Hilbert series, we can read off the quaternionic dimension of the twisted Higgs branch
\begin{equation}
    \dim_\mathbb{H}\mathcal{M}_{\text{tw}}\left[C_{g_1,n}\times C_{g_2}\right]=\begin{dcases}
                N_v=3(g_1-1)+n~, & g_2=1\\
        3(g_2-1)N_v~, & g_2\geq 2~.
    \end{dcases}
\end{equation}
yielding the right-moving central charge
\begin{equation}
    \label{eq:cc-twisted}c_R=6\dim_\mathbb{H}\mathcal{M}_{\text{tw}}\left[C_{g_1,n}\times C_{g_2}\right]=\begin{dcases}
    2n_v~, & g_2=1\\
        6(g_2-1)n_v~, & g_2\geq 2~.
    \end{dcases}
\end{equation}
This result agrees with our conjecture \eqref{tw-g2=1} in the case $g_2=1$ while for $g_2>1$ it matches the anomaly \eqref{cR-tw} associated with $SU(2)_-$. Thus, when $g_2 >1$, it is straightforward to identify the small (0,4) superconformal R-symmetry with $SU(2)_-$. 

On the other hand, as explained when we proposed the conjecture \eqref{tw-g2=1}, the effective 2d theory for $g_2=1$ exhibits enhanced $\mathcal{N}=(4,4)$ supersymmetry. In the infrared, the corresponding small $(4,4)$ superconformal R-symmetry is given by the affine Lie algebra $SU(2)_l \times SU(2)_r$. The left-moving R-symmetry $SU(2)_l$ acts non-trivially on the Cartan components of the gauginos $\lambda_{-,C}$ whereas the right-moving R-symmetry $SU(2)_r$ acts non-trivially on the fermions $\lambda_{+,C}$ in the Cartan parts of the twisted hypermultiplets. The diagonal subgroup of $SU(2)_l \times SU(2)_r$ can then be identified with $SU(2)_-$. From this perspective, each individual factor $SU(2)_l$ or $SU(2)_r$ should be regarded as an \emph{emergent} R-symmetry on the twisted Higgs branch for $g_2=1$: they are invisible in the UV description but manifest only in the infrared. 
With this identification of the $(4,4)$ superconformal R-symmetry, the central charge can be evaluated directly from the infrared spectrum summarized in Table~\ref{tab:fields_on_twisted}. One finds
\be 
c_L=c_R=2n_v~.
\ee

\begin{table}[ht]
\centering    \renewcommand{\arraystretch}{1.3}
\begin{tabular}{c|c|c}
\hline
{$g_2$} & {dim} & {Hilbert series}  \\
\hline
1 & 1 & $\frac{1}{(1-t^2)^2(1+t^2)^{-1}}$ \\
\hline
2 & 3 & $\frac{ 1+3 t^2+t^4}{(1-t^2)^6\left(1+t^2\right)^{-1}}$  \\
\hline
3 & 6 & $\frac{ 1 + 8 t^2 + 14 t^3 + 22 t^4 + 34 t^5 + 22 t^6 + 14 t^7 + 8 t^8 + t^{10}}{(1-t^2)^{12}\left(1+t^2\right)^{-1}}$   \\
\hline
4 & 9 & $\frac{1+17 t^{2}+48 t^{3}+126 t^{4}+320 t^{5}+537 t^{6}+760 t^7+894 t^8+\text{...pal...}+t^{16}}{\left(1-t^2\right)^{18}(1+t^2)^{-1}}$\\
\hline
5 & 12 & $\frac{1+30 t^{2}+110 t^{3}+421 t^{4}+1462 t^{5}+3684 t^{6}+8000 t^{7}+14806 t^{8}+22492 t^{9}+29106 t^{10}+31968 t^{11}+\text{...pal...}+t^{22}}{{\left(1-t^2\right)^{24}}({1+t^2})^{-1}}$\\
\hline
\end{tabular}
\caption{The quaternionic dimensions and contributions to Hilbert series of the twisted Higgs branches from a single gauge node.}
\label{tab:HS_tw}
\end{table}

We note that, for fixed $(g_1,n)$, the number $N_v$ of gauge nodes is
independent of the choice of duality frame.
As a consequence, the Hilbert series of the twisted Higgs branch is
automatically frame-independent.

Additionally, for theories with $n=0$, the Hilbert series of the
twisted Higgs branch is generically \emph{not} symmetric under the exchange
$g_1 \leftrightarrow g_2$, indicating that the exchange symmetry between the two
Riemann surfaces is broken.
As an explicit illustration, 
the Hilbert series of the case $(g_1,g_2)=(2,3)$ takes the form
\begin{equation}
    G_{\text{tw},(2,0)}(t;3)=\left(G_{\text{tw}}(t;3)\right)^3=\left[\frac{ 1 + 8 t^2 + 14 t^3 + 22 t^4 + 34 t^5 + 22 t^6 + 14 t^7 + 8 t^8 + t^{10}}{(1-t^2)^{12}\left(1+t^2\right)^{-1}}\right]^3~,
\end{equation}
while the result for $(g_1,g_2)=(3,2)$ is given by
\begin{equation}
    G_{\text{tw},(3,0)}(t;2)=\left(G_{\text{tw}}(t;2)\right)^6=\left[\frac{ 1+3 t^2+t^4}{(1-t^2)^6\left(1+t^2\right)^{-1}}\right]^6~.
\end{equation}
The clear mismatch between these two expressions provides concrete evidence for the absence of an exchange symmetry between the two Riemann surfaces. 

\section{Outlook}\label{sec:outlook}

In this work, we have proposed conjectural formulas for the right-moving central charges of 2d $\cN=(0,4)$ theories arising from the dimensional reduction of 4d class~$\cS$ theories on a Riemann surface $C_{g_2}$.  
For $G=SU(2)$, we performed detailed checks by analyzing the vacuum moduli spaces and computing Hilbert series of both the special Higgs branches and the twisted Higgs branches.  
These computations confirm our conjectures and reveal several structural features of the resulting theories, which deserve further investigation.

\paragraph{Further study of vacuum moduli spaces.}
We have encountered non-palindromic Hilbert series of the special Higgs branch in cases with $g_1\ge 2$ and $g_2\ge1$. 
This behavior suggests that the geometry of the special Higgs branch undergoes a qualitative structural change in this regime. 
At present, however, we lack a clear physical and geometric interpretation of these non-palindromic cases, and uncovering the underlying structure remains an interesting open problem.

Beyond the two branches analyzed in this work, the remaining irreducible components of the vacuum moduli space also present promising directions for future investigation. 
For instance, in the case $(g_1,g_2,n)=(0,1,5)$, we find 12 additional irreducible components. 
Similarly, for the dumbbell quiver shown in Figure~\ref{fig:reduce_g2n0_dumbbell} with $g_2=1$, there are 5 extra components. 
These examples point to the existence of systematic patterns in the decomposition of the moduli space that are not yet fully understood. 
It would be highly desirable to develop a global description of the full vacuum moduli space that incorporates all such components in a unified framework.

\paragraph{Higher-rank generalizations.}
Although we have focused on $G=SU(2)$ for concreteness, it is clearly desirable to test our central-charge conjectures for general ADE types.  
A natural next step is to extend the Hilbert-series analysis to 2d $(0,4)$ theories with gauge group $G=SU(N)$, $N>2$.  Higher-rank theories generically do not admit Lagrangian descriptions, and it is expected that the interplay between puncture data, flavor symmetry, and the geometry of the special and twisted Higgs branches may give rise to novel phenomena that do not appear in the $SU(2)$ case.

One systematic approach is to start from a torus reduction ($g_2=1$) of a class $\mathcal{S}$ theory of higher rank, which yields a 2d theory with $\mathcal{N}=(4,4)$ supersymmetry. A useful entry point is provided by the $T_N$ theory: the 3d $\mathcal{N}=4$ mirror of its circle reduction has been constructed via brane methods in \cite{Benini:2010uu}, and a further T-duality along the remaining circle should produce the 2d $\mathcal{N}=(4,4)$ theory of interest. Analyzing the vacuum moduli space directly from this brane picture would be illuminating. In particular, it would be interesting to clarify how 3d $\mathcal{N}=4$ mirror symmetry is inherited—or deformed—upon reduction to two dimensions, and what role the resulting 2d $\mathcal{N}=(4,4)$ duality \cite{Diaconescu:1997gu} plays in understanding the geometry of the vacuum moduli spaces.

\paragraph{Magnetic-quiver approach.}
One promising direction is the \emph{magnetic quiver} program (starting from~\cite{Cabrera:2018jxt}), which describes hyper-Kähler Higgs branches of supersymmetric theories in terms of Coulomb branches of 3d $\cN=4$ theories.  
While magnetic quivers have been successfully applied in various dimensions, a systematic formulation for 2d $(0,4)$ theories has not yet been developed.  
Establishing such a framework could provide a powerful tool for identifying moduli spaces of higher-rank theories and for interpreting the structure of special and twisted Higgs branches from a unified perspective.

\paragraph{Brane constructions and string dualities.}
It would also be highly valuable to study the geometry of the special and twisted Higgs branches using brane setups or string dualities.  
Such geometric realizations may offer a more intrinsic understanding of the moduli spaces and may provide guidance for constructing magnetic quivers for $(0,4)$ theories.

\paragraph{Adding punctures on $C_{g_2}$.}
In this work, we restricted attention to reductions on unpunctured $C_{g_2}$.  
Introducing punctures on the second Riemann surface, i.e.\ considering $C_{g_2,n_2}$ with $n_2\geq1$, constitutes an obvious next extension.  
The resulting 2d $(0,4)$ theories are expected to exhibit richer flavor symmetries and more complicated Higgs-branch geometries, and it would be interesting to study how the central charges, 't~Hooft anomalies, and Hilbert series behave in the presence of such punctures.

\paragraph{Toward a 2d TQFT structure.}
Compactifications of class~$\cS$ theories on Riemann surfaces are well known to assemble into a 2d TQFT structure.  
The frame-independence of the special Higgs branch discovered here strongly suggests the existence of a similar TQFT interpretation on $C_{g_2}$ for the resulting $(0,4)$ theories.  
Determining whether such a structure exists—and, if so, identifying its algebraic data—would provide a conceptual explanation for many of the patterns observed in this work.

\paragraph{Other topological twists.}
Finally, one may consider alternative topological twists when reducing $\cT_G[C_{g_1,n}]$ on $C_{g_2}$.  
The topological twist with $U(1)_R \subset \SU(2)_R$ leads to 2d $(2,2)$ theories while that with $\text{diag}(U(1)_R\times U(1)_r)$ produces 2d $(0,2)$ theories~\cite{Gadde:2015wta,Cecotti:2015lab}.  
For genus $g_1>0$, unbroken gauge sectors again appear in the infrared~\cite{Nawata:2023aoq,Jiang:2024ifv}, and it would be interesting to revisit the computation of central charges and moduli spaces in these settings.  
Understanding how different twists modify the IR geometry and the pattern of gapped Abelian sectors may also shed light on the origin of our central-charge formulas.

\paragraph{M5-branes on a more general 4-manifold.}

The compactification of M5-branes on a general 4-manifold~\cite{Gadde:2013sca,Feigin:2018bkf} remains a largely unexplored direction, particularly when the 4-manifold is not simply connected. In such cases, both the structure of the resulting effective 2d theories and their relation to the Vafa--Witten partition functions of 4-manifolds~\cite{Vafa:1994tf,Dijkgraaf:1997ce} are poorly understood. A systematic analysis of these compactifications is still missing and calls for further investigation.

Moreover, as already observed in this work and in the literature~\cite{Benini:2010uu,Gukov:2017zao}, the \emph{order} in which the compactifications of the M5-branes are performed can affect the resulting physics in 2d when the 4-manifold is a product space. Understanding the origin of this phenomenon, and determining when different orders of compactification lead to equivalent or inequivalent 2d theories, would shed new light on the geometry of 6d theories and the nature of their lower-dimensional reductions.

\medskip

In summary, the results obtained here represent only the first step in a broader study of 2d theories originating from class~$\cS$. Moreover, the study of M5-branes on general 4-manifolds represents a challenging and wide-open avenue for future research. 
We hope that the tools and observations developed in this work will serve as a foundation for further progress in understanding their moduli spaces, dualities, and protected observables.

\acknowledgments
The authors would like to thank Noppadol Mekareeya for his guidance on using \texttt{Macaulay2}  for Hilbert series computations. They are especially grateful to Zhenghao Zhong for his valuable insights and for elucidating the conceptual meaning of the Hilbert series, without which this work would have been impossible. They thank Jiahao Zheng for his help and suggestions regarding the  Molien--Weyl calculations.

This work is supported by the Shanghai Municipal Science and Technology Major Project (No.24ZR1403900).

\appendix

\section{Notations}\label{app:notations}

In this appendix, we fix the notational conventions for the description of 4d and 2d fields, as well as for the presentation of their Lagrangians.
In particular, we explain how to write superpotential and $J$-term couplings for theories with gauge group $SU(2)$ in an index-explicit manner.
These conventions will be used systematically to derive the vacuum equations and to compute Hilbert series.

\paragraph{Superpotentials.}
For generic quiver theories, it is often convenient to formulate the Lagrangian in terms of half-hypermultiplets.
We adopt the following basis for the half-hypermultiplets:
\begin{equation}
    Q_{ab1}\equiv Q_{ab}~,\quad Q_{ab2}\equiv \epsilon_{aa^\prime}\epsilon_{bb^\prime}\tilde Q^{a^\prime b^\prime}~,
\end{equation}
where lower indices transform in the fundamental representation of $SU(2)$, while upper indices transform in the anti-fundamental representation.
With this notation, the 4d superpotential term arising from a full hypermultiplet $Q$ coupled to a single $SU(2)$ adjoint multiplet $\Phi$ can be written in the standard form
\begin{equation}
    \tilde Q \Phi Q\equiv \tilde Q^{ac} \Phi^{~b}_{a} Q_{bc}
\end{equation}
or, equivalently, in terms of half-hypermultiplets as
\begin{equation}
    Q_{abc}Q_{a^\prime b^\prime c^\prime}\Phi^{aa^\prime}\epsilon^{bb^\prime}\epsilon^{cc^\prime}~,
\end{equation}
up to an overall numerical constant.
The equivalence of the two expressions can be verified explicitly:
\begin{equation}
\begin{aligned}
    Q_{abc}Q_{a^\prime b^\prime c^\prime}\Phi^{aa^\prime}\epsilon^{bb^\prime}\epsilon^{cc^\prime}&=Q_{ab1}Q_{a^\prime b^\prime 2}\Phi^{aa^\prime}\epsilon^{bb^\prime}-Q_{ab2}Q_{a^\prime b^\prime 1}\Phi^{aa^\prime}\epsilon^{bb^\prime}\\
    &=\epsilon_{a^\prime a^{\prime\prime}}\epsilon_{b^\prime b^{\prime\prime}}\tilde Q^{a^{\prime\prime} b^{\prime\prime}}\Phi^{a^\prime a}Q_{ab}\epsilon^{bb^\prime}-\epsilon_{a a^{\prime\prime}}\epsilon_{b b^{\prime\prime}}\tilde Q^{a^{\prime\prime}b^{\prime\prime}}\Phi^{aa^\prime}Q_{a^\prime b^\prime }\epsilon^{bb^\prime}\\
    &=\epsilon_{a^\prime a^{\prime\prime}}\tilde Q^{a^{\prime\prime} b}\Phi^{a^\prime a}Q_{ab}+\epsilon_{a a^{\prime\prime}}\tilde Q^{a^{\prime\prime}b^\prime}\Phi^{aa^\prime}Q_{a^\prime b^\prime }\\
    &=-2\tilde Q^{ac} \Phi^{~b}_{a} Q_{bc}~,
\end{aligned}
\end{equation}
where in the first equality of the second line, we have used the fact that $\Phi^{aa^\prime}$ is symmetric, as a result of the tracelessness of $\Phi_{a}^{~a^\prime}$.

In a 2d $\mathcal{N}=(0,4)$ theory, an analogous rewriting applies to the $J$-terms.
Using full multiplets, the $J$-terms take the form
\begin{equation}
    \tilde Q^{ac}\tilde \Sigma_{a}^{~b}\Gamma_{bc}+\tilde \Gamma^{ac} \tilde \Sigma_{a}^{~b}Q_{bc}
\end{equation}
Equivalently, in terms of half-multiplets, they can be written as
\begin{equation}
    \Gamma_{abc}Q_{a^\prime b^\prime c^\prime}\tilde \Sigma^{aa^\prime} \epsilon^{bb^\prime}\epsilon^{cc^\prime}~,
\end{equation}
up to an overall sign.
Here $(\Gamma,\tilde\Gamma)$ form a $(0,4)$ Fermi multiplet,
$(Q,\tilde Q)$ form a $(0,4)$ hypermultiplet,
and $(\Sigma,\tilde\Sigma)$ form a $(0,4)$ twisted hypermultiplet.
We denote the corresponding half-multiplets by
\begin{equation}
    Q_{ab1}\equiv Q_{ab}~,\quad Q_{ab2}\equiv \epsilon_{aa^\prime}\epsilon_{bb^\prime}\tilde Q^{a^\prime b^\prime}~,\quad
    \Gamma_{ab1}\equiv \Gamma_{ab}~,\quad\Gamma_{ab2}\equiv \epsilon_{aa^\prime}\epsilon_{bb^\prime}\tilde\Gamma^{a^\prime b^\prime}~.
\end{equation}

\paragraph{$SU(2)$ adjoint representation.}
For $SU(2)$ adjoint multiplets such as a twisted hypermultiplet $\Sigma$, we define
\begin{equation}
    \Sigma^{a a'} \equiv \epsilon^{a b} \, \Sigma_{b}^{~a'} \, ,
\end{equation}
which makes contractions with fundamental multiplets explicit.
We further expand the adjoint fields as
\begin{equation}
    \Sigma^{a a'} = \Sigma^A (e_A)^{a a'}
    \;\equiv\;
    \Sigma^A \, \epsilon^{a b} (e_A)_b^{~a'}~,
    \qquad A = 1,2,3~,
\end{equation}
using the $SU(2)$-triple
\begin{equation}\label{canonical_basis}
    (e_1)_a^{~b}=\begin{pmatrix}
        0 & 1 \\ 0 & 0
    \end{pmatrix}~,\quad (e_2)_a^{~b}=\begin{pmatrix}
        0 & 0 \\ 1 & 0
    \end{pmatrix}~,\quad (e_3)_a^{~b}=\begin{pmatrix}
        1 & 0 \\ 0 & -1
    \end{pmatrix}~,
\end{equation}
which satisfies 
\begin{equation}
    [e_1,e_2]=e_3~,\quad [e_3,e_1]=2e_1~,\quad[e_3,e_2]=-2e_2~.
\end{equation}

\section{Higgs mechanism for \texorpdfstring{$G=SU(2)$}{G=SU(2)}}\label{app:Higgs}
In this appendix, we verify the Higgs mechanism for the special and twisted Higgs branches discussed in Section~\ref{subsec:conjecture} by explicitly computing the mass matrices of gauginos and matter fermions in the case where the gauge groups are $SU(2)$. Due to the local nature of the Higgs mechanism, it is sufficient to analyze a single $SU(2)$ gauge node. The Higgsing of such a node amounts to two basic operations:
\begin{itemize}
    \item Higgsing by (two) hypermultiplets in the fundamental representation of $SU(2)$;
    \item Higgsing by a hypermultiplet in the adjoint representation of $SU(2)$.
\end{itemize}
By examining the Yukawa couplings and explicitly constructing the corresponding fermion mass matrices in these two cases, we demonstrate how chiral fermions become massive when vacuum expectation values are turned on for the scalars in the (twisted) hypermultiplets. This provides a direct and explicit verification of the Higgs mechanism, and therefore the number of massless spectrum, on the special and twisted Higgs branches.

\subsection{Higgsing by fundamental hypermultiplets}
\begin{figure}[ht!]
    \centering
\includegraphics[width=0.35\linewidth]{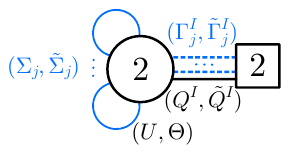}
    \caption{The basic block with hypermultiplets in fundamental representation of both the gauge and flavor $SU(2)$'s.}
    \label{fig:aB1}
\end{figure}

In this subsection, we study the basic block with respect to a fundamental hypermultiplet. As depicted in Figure \ref{fig:aB1}, the 2d $\cN=(0,4)$ field contents are:  one vector multiplet $(U,\Theta)$, two fundamental hypermultiplets $(Q^{I},\tilde{Q}^{I})$, $g_2$ adjoint twisted hypermultiplets $(\Sigma_{j},\tilde{\Sigma}_{j})$, and $2g_2$ fundamental Fermi multiplets $(\Gamma^I_{j},\tilde{\Gamma}^I_{j})$, with $I=1,2$ and $j=1,\ldots, g_2$.
The notation of these superfields is as defined in Table \ref{tab:(0,4)-fields}.

From these field contents, the Lagrangian is constructed as  
\begin{equation}\label{eq:LagFund}
\begin{aligned}
         \mathcal{L} & =  \mathcal{L}_\text{vector} + \mathcal{L}_\text{hyper}+  \mathcal{L}_\text{twist}+  \mathcal{L}_\text{Fermi}+ \mathcal{L}_{J} \\
     & = \int d\theta^+d\overline\theta^+\bigg[~\frac{4}{e^2}\tr(\overline\Upsilon_-\Upsilon_-) +\frac{2}{e^2}\tr(\overline\Theta_-\Theta_- )
     + \sum_{I=1}^{2}~\left( 2\ii\overline Q^I \Dcl_-Q^I-2\ii\tilde Q^I \Dcl_-\overline{\tilde{Q}}^I\right)\\
     &\qquad\qquad\qquad+ \frac{2}{e^2}\sum_{j=1}^{g_2} \tr(-2\ii\overline\Sigma_j \Dcl_-\Sigma_j+2\ii\tilde\Sigma_j \Dcl_-\overline{\tilde{\Sigma}}_j)+\sum_{I=1}^{2}\sum_{j=1}^{g_2} \left(\overline\Gamma^I_{j,-}\Gamma^I_{j,-}-\tilde{\Gamma}^I_{j,-}\overline{\tilde{\Gamma}}^I_{j,-}\right)\bigg] \\
     &\quad+\int d\theta^+~\bigg[\sum_{I=1}^{2}\sum_{j=1}^{g_2} \left(\Gamma^I_{j,-}J_{\Gamma^I_{j}}+\tilde{\Gamma}^I_{j,-}J_{\tilde\Gamma^I_{j}}\right)+\Theta_-J_{\Theta_-}+\mathrm{h.c.}\bigg]~.\\
\end{aligned}
\end{equation}
Here, $ \mathcal{D}_-$ is the covariant derivative, $\Upsilon_-$ is the field strength supermultiplet associated with $U$, and $e$ is the gauge coupling. The structure of interactions is encoded in the $J$- and $E$-term holomorphic functions associated with each Fermi multiplet:  
\begin{equation} \label{eq:appB-EJ1}
J_{\Theta}=-\sum_{I=1}^2\tilde{Q}^I T^A Q^I,\quad  E_\Theta = \sqrt{2}\sum_{j=1}^{g_2}\left[ \Sigma_j, \tilde{\Sigma}_j \right]
\end{equation}
for the Fermi multiplet $\Theta$, and
\begin{equation}
\label{eq:appB-EJ2}
    J_{\Gamma^I_{j}} = -\tilde{Q}^I\tilde{\Sigma}_j  \ ,\quad J_{\tilde{\Gamma}^I_{j}} = -\tilde{\Sigma}_j Q^I,\quad  E_{\Gamma^I_{j}} = \sqrt{2}\Sigma_j Q^I \ ,  \quad E_{\tilde{\Gamma}^I_{j}} = -\sqrt{2}\tilde{Q}^I \Sigma _j
\end{equation}
for Fermi multiplets $\Gamma^I_{j}$ and $\tilde{\Gamma}^I_{j}$.
These terms satisfy the supersymmetry condition ${E} \cdot {J} = 0$~.

\subsubsection*{Special Higgs branch}

For simplicity, we analyze the case where $g_2=1$.
Using the components listed in Table \ref{tab:(0,4)-fields}, the Yukawa terms in the Lagrangian are 
\begin{equation}
\begin{aligned}
\mathcal{L}_{\text{hyper}}&\supset-\sqrt{2}\ii\left(\overline q\lambda_-\psi_++\tilde\psi_+\lambda_-\overline{\tilde q}\right)+\mathrm{h.c.}~,\\
\mathcal{L}_{\text{Fermi}}&\supset-\sqrt{2}\ii\left(-\overline q\lambda_+\psi_--\tilde{\psi}_-\lambda_+\overline{\tilde{q}}\right)+\mathrm{h.c.}~,\\
\mathcal{L}_J&\supset\sqrt{2}\left(\tilde{q}\tilde{\lambda}_-\psi_+ -\tilde{\psi}_+\tilde{\lambda}_-q -\tilde{q}\tilde{\lambda}_+\psi_- +\tilde{\psi}_-\tilde{\lambda}_+q\right)+\mathrm{h.c.}~.
\end{aligned}
\end{equation}
Taking the scalar VEVs $\langle q^I\rangle=v^I,~\langle \tilde{q}^I\rangle=\tilde{v}^I,$
the resulting fermion bilinear terms can be organized into a mass matrix form
\begin{equation} \label{eq:massterms1}
-R_\psi^{\mathsf{T}}M_{\psi_+\lambda_-}L_\lambda-R_\lambda^{\mathsf T}M_{\lambda_+\psi_-}L_\psi+\mathrm{h.c.}~,
\end{equation}
with the mass matrices
\begin{equation}
M_{\lambda_+\psi_-}= M_{\psi_+\lambda_-}^{\mathsf T}~,\qquad M_{\psi_+\lambda_-}
=\sqrt{2}\begin{pmatrix}
    -\ii (v^1_2)^* & (v^1_2)^* & -\ii (v^1_1)^* & (\tilde{v}^1_2) & \ii(\tilde{v}^1_2) & (\tilde{v}^1_1) \\
    -\ii (v^1_1)^* & -(v^1_1)^* & \ii (v^1_2)^* & (\tilde{v}^1_1) & -\ii(\tilde{v}^1_1) & -(\tilde{v}^1_2) \\   
    \ii(\tilde{v}^1_2)^* & (\tilde{v}^1_2)^* & \ii(\tilde{v}^1_1)^* & (v^1_2) & -\ii (v^1_2) & (v^1_1)\\
    \ii(\tilde{v}^1_1)^* & -(\tilde{v}^1_1)^* & -\ii(\tilde{v}^1_2)^* & (v^1_1) & \ii (v^1_1) & -(v^1_2)\\
    -\ii (v^2_2)^* & (v^2_2)^* & -\ii (v^2_1)^* & (\tilde{v}^2_2) & \ii(\tilde{v}^2_2) & (\tilde{v}^2_1) \\
    -\ii (v^2_1)^* & -(v^2_1)^* & \ii (v^2_2)^* & (\tilde{v}^2_1) & -\ii(\tilde{v}^2_1) & -(\tilde{v}^2_2) \\   
    \ii(\tilde{v}^2_2)^* & (\tilde{v}^2_2)^* & \ii(\tilde{v}^2_1)^* & (v^2_2) & -\ii (v^2_2) & (v^2_1)\\
    \ii(\tilde{v}^2_1)^* & -(\tilde{v}^2_1)^* & -\ii(\tilde{v}^2_2)^* & (v^2_1) & \ii (v^2_1) & -(v^2_2)
\end{pmatrix}~,
\end{equation}
and the right- and left-moving basis vectors 
\begin{equation} \label{eq:massbasis1}
R_{\psi} = \begin{pmatrix} \psi^1_{a,+} \\ \tilde{\psi}^1_{a,+} \\  \psi^2_{a,+} \\ \tilde{\psi}^2_{a,+} \\ \end{pmatrix}~, \quad 
L_{\lambda} = \begin{pmatrix} \lambda^A_- \\ \tilde{\lambda}^A_- \\ \end{pmatrix}~,\quad 
R_{\lambda} = \begin{pmatrix} \lambda^A_+ \\ \tilde{\lambda}^A_+ \\ \end{pmatrix}~, \quad
L_{\psi} = \begin{pmatrix} \psi^1_{a,-} \\ \tilde{\psi}^1_{a,-} \\  \psi^2_{a,-} \\ \tilde{\psi}^2_{a,-} \\  \end{pmatrix}~.
\end{equation}
Here the superscripts $I=1,2$ of $\psi_\pm^{I}$ are the flavor index, $a=1,2$ are the fundamental gauge index, and $A=1,2,3$ are the adjoint gauge index. 
From the structure of the mass terms~\eqref{eq:massterms1}, we explicitly observe the pairing between the left-moving gauginos $\lambda_-$ and the right-moving hypermultiplet fermions $\psi_+$, precisely as expected from the Higgs mechanism on the Higgs branch discussed in Section~\ref{subsec:conjecture}.

The fermion masses are obtained by performing a singular value decomposition of the matrix $M_{\psi_+\lambda_-}$. The physical masses are given by the singular values, namely the square roots of the eigenvalues of $M^\dagger M$. Since
$M_{\lambda_+\psi_-}$ is related to $M_{\psi_+\lambda_-}$ by transposition, the two matrices share an identical mass spectrum.

We impose the vacuum equation $J_\Theta=0$ in \eqref{eq:appB-EJ1} for the VEVs $v^I,\tilde v^I,~(I=1,2)$: 
\be 
\tilde v^1T^Av^1+\tilde v^2T^Av^2=0~.
\ee 
Under this condition, both mass matrices $M_{\psi_+ \lambda_-}$ and $M_{\lambda_+ \psi_-}$ exhibit three distinct sets of non-zero singular values, each set being two-fold degenerate.
This structure follows from the block-diagonal form of the corresponding $M^\dagger M$ matrix, yielding
\begin{equation}
m_{1,2}=\sqrt{2K}~, \quad m_{3,4,5,6}=\sqrt{2(K \pm |\vec{D}|)}~,
\end{equation}
where $K$ is the sum of VEVs, and $|\vec{D}|$ is the magnitude of the $D$-term vector $\vec{D} = (D^1, D^2, D^3)$ subject to
\begin{equation}
K = \sum_{I=1}^{2} \left( |v^I|^2 + |\tilde{v}^I|^2 \right)~,\qquad
D^A = \sum_{I=1}^{2} \left( v^{I\dagger} T^A v^I - \tilde{v}^I T^A \tilde{v}^{I\dagger} \right)\equiv0~.
\end{equation}
Since the $D$-terms vanish, all singular values indeed coincide.
We therefore obtain a total of six identical non-zero singular values.

From the mass matrix $M_{\psi_+\lambda_-}$, all six left-moving gauginos $(\lambda^A_-,\tilde{\lambda}^A_-)$ pair with six right-moving hypermultiplet fermions $(\psi^I_{a,+},\tilde{\psi}^I_{a,+})$ and become massive. By supersymmetry, the entire vector multiplet acquires a mass, and the $SU(2)$ gauge symmetry is completely Higgsed.

Similarly, from the mass matrix $M_{\lambda_+\psi_-}$, the right-moving fermions $(\lambda^A_+,\tilde{\lambda}^A_+)$ originating from twisted hypermultiplets pair with an equal number of left-moving fermions from Fermi multiplets $(\psi^I_{a,-},\tilde{\psi}^I_{a,-})$ to become massive. As a result, the twisted hypermultiplets become massive together with the corresponding Fermi multiplets.

\subsubsection*{Twisted Higgs branch}
Let us then consider Higgsing on the twisted Higgs branch for $g_2=1$. In this case, the relevant Yukawa terms from the Lagrangian \eqref{eq:LagFund} are:
\begin{equation}\label{thfmYukawa}
\begin{aligned}
\mathcal{L}_{\text{vector}}&\supset-\frac{2\sqrt{2}}{e^2}\tr\left(
\overline{\tilde{\lambda}}_-[\sigma,{\tilde{\lambda}}_+]+\ii \overline{\tilde{\lambda}}_-[\overline\lambda_+,{\tilde{\sigma}}]
\right)
+\mathrm{h.c.}~,\\
\mathcal{L}_{\text{twist}}&\supset\frac{2\sqrt{2}\ii}{e^2}\tr\left(
[{\tilde{\lambda}}_+,\lambda_-]\overline{\tilde{\sigma}}+\ii\overline\sigma[\lambda_-,\overline\lambda_+]\right)+\mathrm{h.c.}~,\\
\mathcal{L}_{\text{Fermi}}&\supset-\sqrt{2}\left(
\overline\psi_+\overline\sigma\psi_-+\tilde{\psi}_+\sigma\overline{\tilde{\psi}}_-+\mathrm{h.c.}
\right)~,\\
\mathcal{L}_J&\supset\sqrt{2}\left(
\overline{\psi}_+\overline{\tilde{\sigma}}\overline{\tilde{\psi}}_--\tilde{\psi}_+\tilde{\sigma}\psi_-+\mathrm{h.c.}
\right)~.
\end{aligned}
\end{equation}
We construct the fermion mass matrices by turning on vacuum expectation values for the twisted hypermultiplet scalars,
$\langle \sigma^A \rangle \equiv w^A$,
$\langle \tilde{\sigma}^A \rangle \equiv \tilde{w}^A$.
The resulting fermion bilinear terms can be written as
\begin{equation}
-{R'_{\lambda}}^{\mathsf{T}} M_{\lambda_+\lambda_-}L'_{\lambda}-{R'_{\psi}}^{\mathsf{T}}M_{\psi_+\psi_-}L'_{\psi}+\mathrm{h.c.}~,
\end{equation}
where we choose the right- and left-moving bases
\begin{equation}\label{eq:RLbasis2}
    R'_{\lambda} = \begin{pmatrix} \overline{\lambda}^A_+ \\ \tilde{\lambda}^A_+ \\ \end{pmatrix}~, \quad 
L'_{\lambda} = \begin{pmatrix} \lambda^A_- \\ \overline{\tilde{\lambda}}^A_- \\ \end{pmatrix}~,\quad 
R'_{\psi} = \begin{pmatrix} \overline{\psi}^1_{a,+} \\ \tilde{\psi}^1_{a,+} \\  \overline{\psi}^2_{a,+} \\ \tilde{\psi}^2_{a,+} \\ \end{pmatrix}~, \quad 
L'_{\psi} = \begin{pmatrix} \psi^1_{a,-} \\ \overline{\tilde{\psi}}^1_{a,-} \\  \psi^2_{a,-} \\ \overline{\tilde{\psi}}^2_{a,-} \\  \end{pmatrix}~.
\end{equation} 
The mass matrix $M_{\lambda_+\lambda_-}$ encodes the pairing between twisted hypermultiplet fermions $(\lambda_+,\tilde{\lambda}_+)$ and gauginos $(\lambda_-,\tilde{\lambda}_-)$, while $M_{\psi_+\psi_-}$ captures the pairing between hypermultiplet fermions $(\psi_+,\tilde{\psi}_+)$ and Fermi multiplet fermions $(\psi_-,\tilde{\psi}_-)$. This structure precisely realizes the Higgs mechanism on the twisted Higgs branch described in Section~\ref{subsec:conjecture}.

We first analyze the (gaugino--twisted hypermultiplet) mass matrix $M_{\lambda_+\lambda_-}$. In the basis~\eqref{eq:RLbasis2}, it is given explicitly by
\begin{equation} \label{eq:m2tw}
    M_{\lambda_+\lambda_-} = \frac{8\sqrt{2}}{e^2} \begin{pmatrix}
0 & \ii w_3^* & -\ii w_2^* & 0 & -\tilde{w}_3 & \tilde{w}_2 \\
-\ii w_3^* & 0 & \ii w_1^* & \tilde{w}_3 & 0 & -\tilde{w}_1 \\
\ii w_2^* & -\ii w_1^* & 0 & -\tilde{w}_2 & \tilde{w}_1 & 0 \\
0 & \tilde{w}_3^* & -\tilde{w}_2^* & 0 & -\ii w_3 & \ii w_2 \\
-\tilde{w}_3^* & 0 & \tilde{w}_1^* & \ii w_3 & 0 & -\ii w_1 \\
\tilde{w}_2^* & -\tilde{w}_1^* & 0 & -\ii w_2 & \ii w_1 & 0
\end{pmatrix}~,
\end{equation}
The vacuum equation~\eqref{eq:appB-EJ1} implies that a generic solution takes values in the Cartan subalgebra, so that only $w_3$ and $\tilde{w}_3$ are non-vanishing. The singular values of $M_{\lambda_+\lambda_-}$ then yield the fermion masses
\begin{equation} \label{eq:twmass}
    m_{1,2,3,4} = \frac{8\sqrt{2}}{e^2}\sqrt{|w_3|^2+|\tilde{w_3}|^2}~, \qquad m_{5,6}=0~.
\end{equation}
This shows that, on the twisted Higgs branch with $g_2=1$, an $SU(2)$ gauge node coupled to fundamental hypermultiplets retains an unbroken $U(1)$ gauge symmetry.

For $g_2>1$, there are $g_2$ twisted hypermultiplets with VEVs $(w_j,\tilde{w}_j)$, $j=1,\ldots,g_2$. The corresponding mass matrix is a $6\times 6g_2$ matrix composed of $g_2$ identical $6\times6$ blocks of the form~\eqref{eq:m2tw}. The zero modes found above persist only if all vectors $w_j$ and $\tilde{w}_j$ are mutually aligned. However, on the twisted Higgs branch, the $E$-term constraint enforces
\[
\sum_{j=1}^{g_2}\left[w_j,\tilde{w}_j\right] = 0~,
\]
which does \emph{not} require individual commutators such as $[w_i,w_j]$ or $[w_i,\tilde{w}_j]$ to vanish. Consequently, for $g_2>1$, the scalar VEVs generically fail to commute, and the mass matrix becomes full rank, so all gauge symmetry is completely broken.

Next, we consider the (hypermultiplet--Fermi) mass matrix $M_{\psi_+\psi_-}$. In the basis~\eqref{eq:RLbasis2}, it takes the block-diagonal form
\begin{equation}
 M_{\psi_+\psi_-} =  \begin{pmatrix}
M & 0\\
0 & M
\end{pmatrix}~,\qquad 
    M = \sqrt{2} \begin{pmatrix}
w_3^* & w_1^*-\ii w_2^* & -\tilde{w}_3^* & -\tilde{w}_1^*+\ii\tilde{w}_2^* \\
w_1^*+\ii w_2^* & -w_3^* & -\tilde{w}_1^*-\ii\tilde{w}_2^* & \tilde{w}_3^* \\
\tilde{w}_3 & \tilde{w}_1-\ii\tilde{w}_2 & w_3 & w_1-\ii w_2 \\
\tilde{w}_1+\ii\tilde{w}_2 & -\tilde{w}_3 & w_1+\ii w_2 & -w_3 \\
\end{pmatrix}~,
\end{equation}
The VEVs $w$ and $\tilde{w}$ must again satisfy the $E$-term equation~\eqref{eq:appB-EJ1}. For $g_2=1$, the generic solution lies in the Cartan subalgebra, $w=w_3$ and $\tilde{w}=\tilde{w}_3$, yielding
\[
\det M = |w_3|^4 + 2|w_3|^2|\tilde{w}_3|^2 + |\tilde{w}_3|^4 \neq 0~.
\]
Thus, all singular values of $M$ are generically non-zero, and all hypermultiplet fermions pair with Fermi multiplet fermions to become massive.

\subsection{Higgsing by an adjoint hypermultiplet}

\begin{figure}[ht!]
    \centering
    \includegraphics[width=0.4\linewidth]{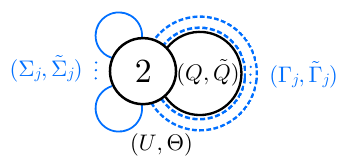}
    \caption{Gauge node with adjoint hypermultiplet}
    \label{fig:aB2}
\end{figure}

We now turn to the Higgs mechanism induced by an adjoint hypermultiplet.
As illustrated in Figure~\ref{fig:aB2}, the 2d (0,4) field content consists of a single vector multiplet $(U,\Theta)$, one adjoint hypermultiplet $(Q,\tilde{Q})$, $g_2$ adjoint twisted hypermultiplets $(\Sigma_j,\tilde{\Sigma}_j)$, and $g_2$ adjoint Fermi multiplets $(\Gamma_j,\tilde{\Gamma}_j)$, with $j=1,\ldots,g_2$.

The Lagrangian takes the same general form as in~\eqref{eq:LagFund}, with the fundamental hypermultiplet and Fermi multiplets replaced by fields transforming in the adjoint representation.
Correspondingly, the $J$- and $E$-term interactions are adjoint analogues of those in~\eqref{eq:appB-EJ1} and~\eqref{eq:appB-EJ2}.

\subsubsection*{Special Higgs Branch}
We focus on the Yukawa interaction terms relevant to the Special Higgs branch:
\begin{equation}
\begin{aligned}
\mathcal{L}_{\text{hyper}} &\supset -\frac{2\sqrt{2}\mathrm{i}}{e^2}\mathrm{tr}\left(\overline{q}[\lambda_-,\psi_+]+\tilde{\psi}_+[\lambda_-,\overline{\tilde{q}}]\right)+\mathrm{h.c.}~, \\
\mathcal{L}_{\text{Fermi}} &\supset -\frac{2\sqrt{2}}{e^2}\mathrm{tr}\left(
-\mathrm{i}\overline{q}[\lambda_+,\psi_-]-\mathrm{i}\tilde{\psi}_-[\lambda_+,\overline{\tilde{q}}]
\right)+\mathrm{h.c.}~, \\
\mathcal{L}_J &\supset \frac{2\sqrt{2}}{e^2}\mathrm{tr}
\left(
\tilde{q}[\tilde{\lambda}_-,\psi_+] -\tilde{\psi}_+[\tilde{\lambda}_-,q] -\tilde{q}[\tilde{\lambda}_+,\psi_-] +\tilde{\psi}_-[\tilde{\lambda}_+,q] \right)+\mathrm{h.c.}~.
\end{aligned}
\end{equation}
All fermions transform in the adjoint representation of the $SU(2)$ gauge group. Without loss of generality, we assume $g_2=1$. 
Upon taking the scalar VEVs $\langle q^A\rangle=v^A,~\langle \tilde{q}^A\rangle=\tilde{v}^A$, we organize these fermions in the following basis 
\begin{equation}\label{eq:basis-adj}
  R_{\psi} = \begin{pmatrix} \psi^A_+ \\ \tilde{\psi}^A_+ \end{pmatrix}~, \quad 
L_{\lambda} = \begin{pmatrix} \lambda^A_- \\ \tilde{\lambda}^A_- \end{pmatrix}~, \quad 
R_{\lambda} = \begin{pmatrix} \lambda^A_+ \\ \tilde{\lambda}^A_+ \end{pmatrix}~, \quad 
L_{\psi} = \begin{pmatrix} \psi^A_- \\ \tilde{\psi}^A_- \end{pmatrix}~.
\end{equation}
The fermion mass terms then take the form
\begin{equation}
-R_{\psi}^{\mathsf T}M_{\psi_+\lambda_-}L_{\lambda}-R_{\lambda}^{\mathsf T}M_{\lambda_+\psi_-}L_{\psi}+\mathrm{h.c.}~.
\end{equation}
As before, the mass matrices $M_{\psi_+\lambda_-}$ and $M_{\lambda_+\psi_-}$ encode the pairing pattern between right- and left-moving fermions anticipated in Section~\ref{subsec:conjecture}.

The explicit form of the mass matrix $M_{\psi_+\lambda_-}$ in the basis~\eqref{eq:basis-adj} is 
\begin{equation}
M_{\psi_+\lambda_-} = \frac{8\sqrt{2}}{e^2} \begin{pmatrix}
0 & -v_3^* & v_2^* & 0 & -\mathrm{i}\tilde{v}_3 & \mathrm{i}\tilde{v}_2 \\
v_3^* & 0 & -v_1^* & \mathrm{i}\tilde{v}_3 & 0 & -\mathrm{i}\tilde{v}_1 \\
-v_2^* & v_1^* & 0 & -\mathrm{i}\tilde{v}_2 & \mathrm{i}\tilde{v}_1 & 0 \\
0 & -\tilde{v}_3^* & \tilde{v}_2^* & 0 & \mathrm{i} v_3 & -\mathrm{i} v_2 \\
\tilde{v}_3^* & 0 & -\tilde{v}_1^* & -\mathrm{i} v_3 & 0 & \mathrm{i} v_1 \\
-\tilde{v}_2^* & \tilde{v}_1^* & 0 & \mathrm{i} v_2 & -\mathrm{i} v_1 & 0
\end{pmatrix}~.
\end{equation}
Using the vacuum equation $J_{\Theta}=0$, the physical fermion masses obtained from the singular values are
\begin{equation}\label{eq:mass-adj-sHiggs}
m_{1,2,3,4} = \frac{8\sqrt{2}}{e^2}\sqrt{|v|^2+|\tilde{v}|^2}~, \qquad m_{5,6}=0~.
\end{equation}
We therefore find that the Cartan components of the left-moving gauginos, $\lambda_-^3$ and $\tilde{\lambda}_-^3$, together with their right-moving counterparts, remain massless. Consequently, an unbroken $U(1)$ gauge symmetry survives on the special Higgs branch.

The second mass matrix is related to the first by
$M_{\lambda_+\psi_-} = M_{\psi_+\lambda_-}^{\mathsf T}$, and thus has a mass spectrum identical to that of $M_{\psi_+\lambda_-}$. The corresponding zero modes imply that the Cartan components of the twisted hypermultiplets and Fermi multiplets remain massless, while the remaining components acquire the same masses as~\eqref{eq:mass-adj-sHiggs}.
This confirms the prediction on the special Higgs branch in Section \ref{subsec:conjecture}.

\subsubsection*{Twisted Higgs Branch}
We now turn to the twisted Higgs branch. The relevant Yukawa interaction terms are
\begin{equation}
\begin{aligned}
\mathcal{L}_{\text{vector}} &\supset -\frac{2\sqrt{2}}{e^2}\mathrm{tr}\left(\overline{\tilde{\lambda}}_-[\sigma,{\tilde{\lambda}}_+]+\ii \overline{\tilde{\lambda}}_-[\overline\lambda_+,{\tilde{\sigma}}]\right)+\mathrm{h.c.}~, \\
\mathcal{L}_{\text{twist}} &\supset \frac{2\sqrt{2}\ii}{e^2}\mathrm{tr}\left([{\tilde{\lambda}}_+,\lambda_-]\overline{\tilde{\sigma}}+\ii\overline\sigma[\lambda_-,\overline\lambda_+]\right)+\mathrm{h.c.}~, \\
\mathcal{L}_{\text{Fermi}} &\supset -\frac{2\sqrt{2}}{e^2}\mathrm{tr}\left(\overline\psi_+[\overline\sigma,\psi_-]+\tilde{\psi}_+[\sigma,\overline{\tilde{\psi}}_-]\right)+\mathrm{h.c.}~, \\
\mathcal{L}_J &\supset \frac{2\sqrt{2}}{e^2}\mathrm{tr}\left(\overline{\psi}_+[\overline{\tilde{\sigma}},\overline{\tilde{\psi}}_-]-\tilde{\psi}_+[\tilde{\sigma},\psi_-]\right)+\mathrm{h.c.}~.
\end{aligned}
\end{equation}
We evaluate these terms by turning on vacuum expectation values for the twisted hypermultiplet scalars,
$\langle\sigma^A\rangle \equiv w^A$ and $\langle\tilde{\sigma}^A\rangle \equiv \tilde{w}^A$. The resulting fermion mass terms take the form
\begin{equation}
  -{R'_{\lambda}}^{\mathsf{T}} M_{\lambda_+\lambda_-}L'_{\lambda}  -{{R}'_{\psi}}^{\mathsf T}{M}_{\psi_+\psi_-}{L}'_{\psi}+\mathrm{h.c.}~.
\end{equation}
The first mass term is identical to that of the fundamental case. The basis vectors $R'_{\lambda}$ and $L'_{\lambda}$ are defined in~\eqref{eq:RLbasis2}, and the mass matrix $M_{\lambda_+\lambda_-}$ is given in~\eqref{eq:m2tw}. The analysis therefore parallels that of the fundamental case. For $g_2=1$, a generic solution to the vacuum equations requires $w=w_3$ and $\tilde{w}=\tilde{w}_3$, leading to the fermion mass spectrum~\eqref{eq:twmass}, with two massless modes corresponding to an unbroken $U(1)$ gauge symmetry. For $g_2>1$, a generic choice of VEVs breaks all gauge symmetries completely.

The second mass term deserves further discussion. The corresponding mass matrix is
\[
\mathcal{M}_{\psi_+\psi_-} = \frac{8\sqrt{2}}{e^2} \begin{pmatrix}
0 & \mathrm{i} w_3^* & -\mathrm{i} w_2^* & 0 & -\mathrm{i}\tilde{w}_3^* & \mathrm{i}\tilde{w}_2^* \\
-\mathrm{i} w_3^* & 0 & \mathrm{i} w_1^* & \mathrm{i}\tilde{w}_3^* & 0 & -\mathrm{i}\tilde{w}_1^* \\
\mathrm{i} w_2^* & -\mathrm{i} w_1^* & 0 & -\mathrm{i}\tilde{w}_2^* & \mathrm{i}\tilde{w}_1^* & 0 \\
0 & \mathrm{i}\tilde{w}_3 & -\mathrm{i}\tilde{w}_2 & 0 & \mathrm{i} w_3 & -\mathrm{i} w_2 \\
-\mathrm{i}\tilde{w}_3 & 0 & \mathrm{i}\tilde{w}_1 & -\mathrm{i} w_3 & 0 & \mathrm{i} w_1 \\
\mathrm{i}\tilde{w}_2 & -\mathrm{i}\tilde{w}_1 & 0 & \mathrm{i} w_2 & -\mathrm{i} w_1 & 0
\end{pmatrix}~.
\]
with the basis vectors
\[
{R}'_{\psi} = \begin{pmatrix} \overline{\psi}^A_+ \\ \tilde{\psi}^A_+ \end{pmatrix}~, \quad 
{L}'_{\psi} = \begin{pmatrix} \psi^A_- \\ \overline{\tilde{\psi}}^A_- \end{pmatrix}~.
\]
For $g_2=1$, taking the generic Cartan-valued VEVs $w=w_3$ and $\tilde{w}=\tilde{w}_3$ yields the mass spectrum
\[
m_{1,2,3,4} = \frac{8\sqrt{2}}{e^2}\sqrt{|w|^2+|\tilde{w}|^2}~, \qquad m_{5,6}=0~, 
\]
Thus, the Cartan components of the hypermultiplet fermions can remain massless.
However, this situation occurs only for $(g_1,n)=(1,1)$ with $g_2=1$, where the twisted Higgs branch becomes a subvariety of the special Higgs branch.  
For all other values of $(g_1,n)$, the Higgsing of the remaining gauge nodes lifts these massless hypermultiplet modes, so that they acquire non-zero masses.

For $g_2>1$, a generic choice of vacuum expectation values renders all hypermultiplet fermions massive.

\section{Examples of Hilbert series}\label{app:hilbert-series}

In this appendix, we present explicit examples illustrating the computation of Hilbert series for the Higgs branches discussed in Section~\ref{sec:vac-moduli}. For a variety of choices of $(g_1,n,g_2)$, we first write down the corresponding vacuum equations and then determine the defining ideals of the Higgs branches. Using these data, we compute the associated Hilbert series, which encode the structure of holomorphic functions on the vacuum moduli spaces.

The purpose of this appendix is twofold. First, these explicit computations provide concrete realizations of the general framework developed in the main text and serve as nontrivial checks of the proposed descriptions of the Higgs branches. Second, the explicit evaluation of the Hilbert series allows us to elucidate several structural properties, including their independence of the choice of duality frame, the identification of certain special Higgs branches with nilpotent orbits, and the appearance of non-palindromic Hilbert series.

The last two rows of Tables~\ref{tab:HS_(1,2)} and~\ref{tab:HS_(2,0)} are obtained exclusively using the gluing method. All other computations in this appendix are carried out using standard algebraic--geometric techniques, including the construction of polynomial ideals and the evaluation of Hilbert series with \texttt{Macaulay2}, and are independently verified using the gluing method.

\subsection*{Example: \texorpdfstring{$(g_1,n)=(0,3)$}{(g1,n)=(0,3)}}\label{subsec:(0,3)}
\begin{figure}[ht!]
    \centering
    \includegraphics[width=0.55\linewidth]{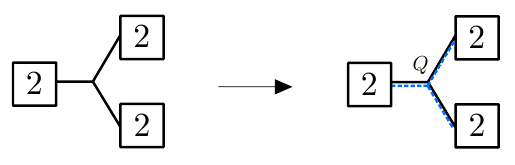}
    \caption{The quiver diagrams of the class $\mathcal{S}$ theory $\mathcal{T}[C_{0,3}]$ and the corresponding 2d (0,4) theory upon the reduction on $C_{g_2}$. For the 2d theory, the solid black lines represent one fundamental hypermultiplet $Q_{abc}$. The dashed blue lines represent $g_2$ fundamental Fermi multiplets $\Gamma_{jabc},~j=1,\cdots g_2$.}
    \label{fig:reduce_g0n3}
\end{figure}

The fundamental building block in our construction is the \emph{trinion theory}. 
The associated 4d and 2d quiver diagrams are shown in Figure~\ref{fig:reduce_g0n3}. 
Since the trinion theory contains no gauge multiplets, there are no $J$-term or $E$-term constraints. 
As a result, the Higgs branch is completely unconstrained and is freely generated by the scalar fields $q_{abc}$.
The corresponding Hilbert series therefore takes a particularly simple form,
\begin{equation}
    G_{(0,3)}(t,x_j;g_2)
    =\frac{1}{1 - t\, x_1^{\pm 1} x_2^{\pm 1} x_3^{\pm 1}}~,
\end{equation}
which counts holomorphic operators built from the $q_{abc}$.
This describes the Higgs branch, whose quaternionic dimension is $4$.
Indeed, the Higgs branch of the trinion theory is simply the flat space $\mathbb{H}^4$.

\subsection*{Example: \texorpdfstring{$(g_1,n)=(0,4)$}{(g1,n)=(0,4)}}\label{subsec:(0,4)}

\begin{figure}[ht!]
    \centering
    \includegraphics[width=0.7\linewidth]{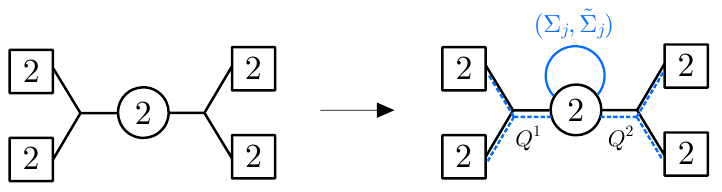}
    \caption{The quiver diagrams of the class $\mathcal{S}$ theory $\mathcal{T}[C_{0,4}]$ and the corresponding 2d (0,4) theory upon the reduction on $C_{g_2}$. For the 2d theory, the solid black lines represent two fundamental hypermultiplets $Q^I_{abc},I=1,2$. The solid blue line represents $g_2$ adjoint twisted hypermultiplets $\Sigma_j,\tilde\Sigma_j$ and the dashed blue lines represent $2g_2$ fundamental Fermi multiplets $\Gamma_{jabc}^I,~I=1,2,~j=1,\cdots g_2$.}
    \label{fig:reduce_g0n4}
\end{figure}
The 4d quiver and the corresponding 2d (0,4) quiver are shown in Figure~\ref{fig:reduce_g0n4}. 
The $J$-term and $E$-term equations take the form
\begin{equation}\label{JE_eqn(0,4)}
    \begin{aligned}
        \left(q^1_{abc} q^1_{a' b' c'} + q^2_{abc} q^2_{a' b' c'}\right)
        (e_A)^{a a'} \, \epsilon^{b b'} \epsilon^{c c'} &= 0~, \\
        q^I_{a' b' c'} \, \sigma_j^{a a'} \, \epsilon^{b b'} \epsilon^{c c'}
        = q^I_{a' b' c'} \, \tilde{\sigma}_j^{a a'} \, \epsilon^{b b'} \epsilon^{c c'} &= 0~, \\
        \sum_{j=1}^{g_2} \bigl[ \sigma_j , \tilde{\sigma}_j \bigr] &= 0~,
    \end{aligned}
\end{equation}
where $e_A$ denote the canonical basis elements of the adjoint representation, as defined in \eqref{canonical_basis}.

The ideal generated by the equations in \eqref{JE_eqn(0,4)} admits two non-trivial minimal prime ideals, corresponding to two distinct irreducible components of the vacuum moduli space: the \emph{special Higgs branch} and the \emph{twisted Higgs branch}.
\begin{itemize}
    \item The special Higgs branch is characterized by the constraints $\sigma_j = \tilde{\sigma}_j = 0$ for all $j$. 
    It has quaternionic dimension $5$. 
    The corresponding Hilbert series is
    \begin{equation}
        G_{H,(0,4)}(t, x_j = 1; g_2)
        = \frac{(1 + t^2)\bigl(1 + 17 t^2 + 48 t^4 + 17 t^6 + t^8\bigr)}
        {(1 - t^2)^{10}}~.
    \end{equation}
    This expression coincides with the Higgs branch Hilbert series of the 4d theory $\mathcal{T}[C_{0,4}]$. 
    Moreover, it agrees with the Hilbert series of the nilpotent orbit of type $D_4$ associated with the partition $(2^2,1^4)$~\cite{Hanany:2016gbz}.

    \item   
    The Hilbert series of the twisted Higgs branch is obtained from the general expression \eqref{eq:tw_HS_general} by setting $N_v = 1$.
\end{itemize}

\subsubsection*{Example: \texorpdfstring{$(g_1,n)=(1,1)$}{(g1,n)=(1,1)}}\label{subsec:(1,1)}

\begin{figure}[ht!]
    \centering
    \includegraphics[width=0.6\linewidth]{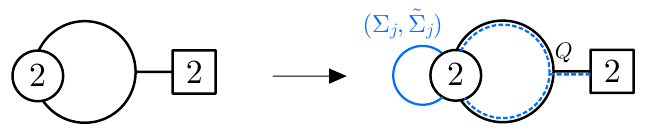}
    \caption{The quiver diagrams of the class $\mathcal{S}$ theory $\mathcal{T}[C_{1,1}]$ and the corresponding 2d (0,4) theory upon the reduction on $C_{g_2}$. For the 2d theory, the solid black lines represent one fundamental hypermultiplet $Q_{abc}$. The solid blue line represents $g_2$ adjoint twisted hypermultiplets $\Sigma_j,\tilde\Sigma_j$ and the dashed blue lines represent $g_2$ fundamental Fermi multiplets $\Gamma_{jabc},~j=1,\cdots g_2$.}
    \label{fig:reduce_g1n1}
\end{figure}

The 4d quiver and the corresponding 2d (0,4) quiver diagrams are shown in Figure \ref{fig:reduce_g1n1}. The $J$-term and $E$-term equations are
\begin{equation}\label{JE_eqn(1,1)}
    \begin{aligned}
        q_{abc}q_{a^\prime b^\prime c^\prime}\left[(e_A)^{aa^\prime}\epsilon^{bb^\prime}\epsilon^{cc^\prime}+\epsilon^{aa^\prime}(e_A)^{bb^\prime}\epsilon^{cc^\prime}\right]&=0~,\\
        q_{a^\prime b^\prime c^\prime}\left(\sigma_j^{aa^\prime}\epsilon^{bb^\prime}\epsilon^{cc^\prime}+\epsilon^{aa^\prime}\sigma_j^{bb^\prime}\epsilon^{cc^\prime}\right)=q_{a^\prime b^\prime c^\prime}\left(\tilde\sigma_j^{aa^\prime}\epsilon^{bb^\prime}\epsilon^{cc^\prime}+\epsilon^{aa^\prime}\tilde\sigma_j^{bb^\prime}\epsilon^{cc^\prime}\right)&=0~,\\
        \sum_{j=1}^{g_2}\left[\sigma_j,\tilde\sigma_j\right]&=0~. 
    \end{aligned}
\end{equation}

The ideal generated by \eqref{JE_eqn(1,1)} has two non-trivial minimal prime ideals corresponding to the special Higgs branch and the twisted Higgs branch, respectively. 
\begin{itemize}
    \item     The special Higgs branch has quaternionic dimension $g_2 + 2$ and is characterized by the constraints
\begin{equation}\label{sH_constrain_(1,1)}
    [\sigma_i,\sigma_j]=[\sigma_i,\tilde\sigma_j]=[\tilde\sigma_i,\tilde\sigma_j]=[\sigma_j,q]=[\sigma_j,\tilde q]=[\tilde\sigma_j,q]=[\tilde\sigma_j,\tilde q]=0~,\quad \forall~i,j=1,\cdots g_2~,
\end{equation}
These conditions imply that $\sigma_j$, $\tilde{\sigma}_j$, $q$, and $\tilde{q}$ are mutually commuting and hence can be simultaneously diagonalized; equivalently, they are all aligned along a common Cartan direction.    The corresponding Hilbert series takes the form
\begin{equation}\label{(1,1)specialhiggs}
    G_{H,(1,1)}(t,x=1;g_2)=\frac{1}{(1-t)^2}\cdot\frac{1+(g_2+1)(2g_2+1)t^2+\cdots}{(1-t^2)^{2g_2+2}}~,
\end{equation}
 where the factor $(1 - t)^{-2}$ accounts for the contribution of two free hypermultiplets arising from the gauge-invariant combinations $\epsilon^{ab} q_{abc}$.

Upon factoring out these free hypermultiplets, the Hilbert series of the remaining factor matches precisely that of the closure of the nilpotent orbit of type $C_{g_2+1}$ associated with the partition $(2,1^{2g_2})$~\cite{Hanany:2016gbz}.
More concretely, the special Higgs branch decomposes as a direct product
\be
\cM_H \;\cong\; \bH \;\times\; \bigg(\frac{\bC^2 \times (\bC^2)^{g_2}}{\bZ_2}\bigg)~.
\ee
The first factor $\bH \cong \bC^2$ is the Higgs branch contribution of the single free hypermultiplet.
The second factor is the closure of the nilpotent orbit of type $C_{g_2+1}$ labelled by the partition $(2,1^{2g_2})$ \cite[App. A]{Bourget:2020asf}, realized geometrically as follows.
The factor $\bC^2$ parametrizes the two complex scalar fields of the adjoint hypermultiplet that lie along the Cartan subalgebra of $\mathfrak{su}(2)$, while $(\bC^2)^{g_2}$ parametrizes the analogous Cartan-valued scalars of the $g_2$ adjoint twisted hypermultiplets.
The residual $\bZ_2$ Weyl group of $SU(2)$ acts simultaneously and diagonally on both $\bC^2$ and $(\bC^2)^{g_2}$, producing the orbifold quotient.

    The explicit unrefined Hilbert series for $g_2 = 0, \ldots, 4$ are listed in Table~\ref{tab:HS_(1,1)}.
\begin{table}[ht]
\centering    \renewcommand{\arraystretch}{1.3}
\begin{tabular}{c|c|c}
\hline
{$g_2$} & {dim} & {Hilbert series}  \\
\hline
0 & 2 & $\frac{1}{(1-t)^2}\frac{1+t^2}{\left(1-t^2\right)^2}$ \\
\hline
1 & 3 & $\frac{1}{(1-t)^2}\frac{1+6 t^2+t^4}{\left(1-t^2\right)^4}$ \\
\hline
2 & 4 & $\frac{1}{(1-t)^2}\frac{(1+t^2) (t^4+14 t^2+1)}{(1-t^2)^6}$  \\
\hline
3 & 5 & $\frac{1}{(1-t)^2}\frac{1+28 t^2+70 t^4+28 t^6+t^8}{\left(1-t^2\right)^8}$   \\
\hline
4 & 6 & $\frac{1}{(1-t)^2}\frac{(1+t^2) (1+44 t^2+166 t^4+44 t^6+t^8)}{\left(1-t^2\right)^{10}}$ \\
\hline
\end{tabular}
\caption{The quaternionic dimensions and unrefined Hilbert series of the special Higgs branches of the case $(g_1,n)=(1,1)$.}
\label{tab:HS_(1,1)}
\end{table}
\item For $g_2=1$, the twisted Higgs branch is a subvariety of the special Higgs branch.  For $g_2 \geq 2$, the Hilbert series is obtained from the general formula \eqref{eq:tw_HS_general} with $N_v = 1$, together with an additional factor of $(1 - t)^{-2}$ accounting for the free hypermultiplets.
\end{itemize}

\subsubsection*{Example: \texorpdfstring{$(g_1,n)=(1,2)$}{(g1,n)=(1,2)}, frame 1}\label{subsec:(1,2)frame1}

\begin{figure}[ht]
    \centering
    \includegraphics[width=0.7\linewidth]{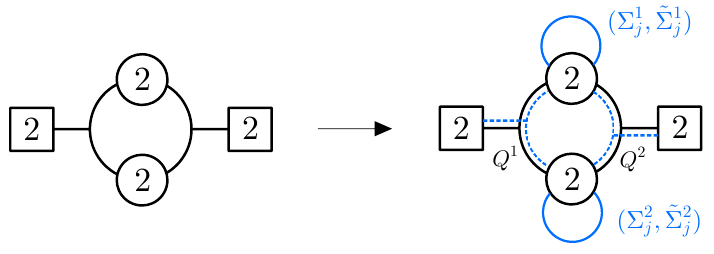}
    \caption{The quiver diagrams of the class $\mathcal{S}$ theory $\mathcal{T}[C_{1,2}]$ and the corresponding 2d (0,4) theory upon the reduction on $C_{g_2}$, in the first frame. For the 2d theory, the solid black lines represent two fundamental hypermultiplets $Q^I_{abc},I=1,2$. The solid blue line represents $2g_2$ adjoint twisted hypermultiplets $\Sigma_j^{J},\tilde\Sigma_j^{J},J=1,2$ and the dashed blue lines represent $2g_2$ fundamental Fermi multiplets $\Gamma^I_{jabc},~I=1,2,~j=1,\cdots g_2$.}
    \label{fig:reduce_g1n2_f1}
\end{figure}

For the case $(g_1,n) = (1,2)$, there are two different frames, shown in
Figures~\ref{fig:reduce_g1n2_f1} and~\ref{fig:reduce_g1n2_f2} (tadpole frame). 
We first analyze the frame depicted in Figure~\ref{fig:reduce_g1n2_f1}. 
The corresponding $J$-term and $E$-term equations are
\begin{equation}\label{Fterm_f1}
    \begin{aligned}
        \left(q^1_{abc}q^1_{a^\prime b^\prime c^\prime}+q^2_{abc}q^2_{a^\prime b^\prime c^\prime}\right)(e_A)^{a\ap}\epsilon^{b\bp}\epsilon^{c\cp}=\left(q^1_{abc}q^1_{a^\prime b^\prime c^\prime}+q^2_{abc}q^2_{a^\prime b^\prime c^\prime}\right)\epsilon^{a\ap}(e_A)^{b\bp}\epsilon^{c\cp}&=0~,\\
        q^I_{\ap\bp\cp}\left(\sigma_{j}^{1a\ap}\epsilon^{b\bp}\epsilon^{c\cp}+\epsilon^{a\ap}\sigma_{j}^{2b\bp}\epsilon^{c\cp}\right)=q^I_{\ap\bp\cp}\left(\tilde\sigma_{j}^{1a\ap}\epsilon^{b\bp}\epsilon^{c\cp}+\epsilon^{a\ap}\tilde\sigma_{j}^{2b\bp}\epsilon^{c\cp}\right)&=0~,\\
        \sum_{j=1}^{g_2}\left[\sigma^J_j,\tilde\sigma^J_j\right]&=0~.
    \end{aligned}
\end{equation}

The ideal generated by \eqref{Fterm_f1} has two non-trivial minimal prime ideals corresponding to the special Higgs branch and the twisted Higgs branch, respectively. 
\begin{table}[ht]
\centering    \renewcommand{\arraystretch}{1.3}
\begin{tabular}{c|c|c}
\hline
{$g_2$} & {dim} & {Hilbert series}  \\
\hline
0 & 3 & $\frac{ 1+3 t^2+t^4}{\left(1-t^2\right)^6(1+t^2)^{-1}}$ \\
\hline
1 & 4 & $\frac{ 1+4 t^2+10 t^3+4 t^4+t^6}{\left(1-t^2\right)^8\left(1+t^2\right)^{-1}}$ \\
\hline
2 & 5 & $\frac{ 1+9 t^2+20 t^3+20 t^4+20 t^5+9 t^6+t^8}{\left(1-t^2\right)^{10}\left(1+t^2\right)^{-1}}$  \\
\hline
3 & 6 & $\frac{ 1+18 t^2+30 t^3+61 t^4+100 t^5+61 t^6+30 t^7+18 t^8+t^{10}}{\left(1-t^2\right)^{12}\left(1+t^2\right)^{-1}}$   \\
\hline
4 & 7 & $\frac{ 1+31 t^{2}+40 t^3+155 t^4+280 t^5+266 t^6+280 t^7+155 t^8+40 t^9+31 t^{10}+t^{12}}{\left(1-t^2\right)^{14}\left(1+t^2\right)^{-1}}$ \\
\hline
\end{tabular}
\caption{The quaternionic dimensions and unrefined Hilbert series of the special Higgs branches of the case $(g_1,n)=(1,2)$.}
\label{tab:HS_(1,2)}
\end{table}

\begin{itemize}
    \item  The special Higgs branch has quaternionic dimension $g_2 + 3$. 
    Its unrefined Hilbert series takes the general form
\begin{equation}
    G_{H,(1,2)}(t,x_j=1;g_2)=\frac{1+\cdots}{(1-t^2)^{2g_2+6}(1+t^2)^{-1}}~.
\end{equation}
    The explicit expressions for $g_2 = 0, \ldots, 4$ are summarized in Table~\ref{tab:HS_(1,2)}.

\item  The Hilbert series of the twisted Higgs branch is given by the general expression
    \eqref{eq:tw_HS_general} with $N_v = 2$.
\end{itemize}

\subsubsection*{Example: \texorpdfstring{$(g_1,n)=(1,2)$}{(g1,n)=(1,2)}, frame 2 (tadpole frame)}\label{subsec:(1,2)frame2}

\begin{figure}[ht]
    \centering
    \includegraphics[width=0.8\linewidth]{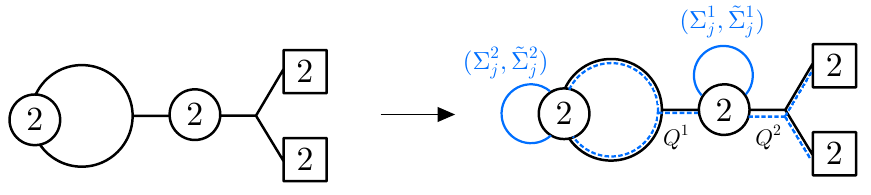}
    \caption{The quiver diagrams of the class $\mathcal{S}$ theory $\mathcal{T}[C_{1,2}]$ and the corresponding 2d (0,4) theory upon the reduction on $C_{g_2}$, in the second frame. For the 2d theory, the solid black lines represent two fundamental hypermultiplets $Q^I_{abc},I=1,2$. The solid blue line represents $2g_2$ adjoint twisted hypermultiplets $\Sigma_j^{J},\tilde\Sigma_j^{J},J=1,2$ and the dashed blue lines represent $2g_2$ fundamental Fermi multiplets $\Gamma^I_{jabc},~I=1,2,~j=1,\cdots g_2$.}
    \label{fig:reduce_g1n2_f2}
\end{figure}

We now turn to the \emph{tadpole frame} for the case $(g_1,n) = (1,2)$, illustrated in Figure~\ref{fig:reduce_g1n2_f2}. 
In this frame, the $J$-term and $E$-term equations are given by
\begin{equation}\label{Fterm_f2}
    \begin{aligned}
        \left(q^1_{abc}q^1_{a^\prime b^\prime c^\prime}+q^2_{abc}q^2_{a^\prime b^\prime c^\prime}\right)(e_A)^{a\ap}\epsilon^{b\bp}\epsilon^{c\cp}=q^1_{abc}q^1_{\ap\bp\cp}\left[\epsilon^{a\ap}(e_A)^{b\bp}\epsilon^{c\cp}+\epsilon^{a\ap}\epsilon^{b\bp}(e_A)^{c\cp}\right]&=0~,\\
        q^1_{\ap\bp\cp}\left(\sigma_{j}^{1a\ap}\epsilon^{b\bp}\epsilon^{c\cp}+\epsilon^{a\ap}\sigma_{j}^{2b\bp}\epsilon^{c\cp}+\epsilon^{a\ap}\epsilon^{2b\bp}\sigma_{j}^{c\cp}\right)=q^2_{\ap\bp\cp}\sigma_{j}^{1a\ap}\epsilon^{b\bp}\epsilon^{c\cp}&=0~,\\
        q^1_{\ap\bp\cp}\left(\tilde\sigma_{j}^{1a\ap}\epsilon^{b\bp}\epsilon^{c\cp}+\epsilon^{a\ap}\tilde\sigma_{j}^{2b\bp}\epsilon^{c\cp}+\epsilon^{a\ap}\epsilon^{2b\bp}\tilde\sigma_{j}^{c\cp}\right)=q^2_{\ap\bp\cp}\tilde\sigma_{j}^{1a\ap}\epsilon^{b\bp}\epsilon^{c\cp}&=0~,\\
        \sum_{j=1}^{g_2}\left[\sigma^J_j,\tilde\sigma^J_j\right]&=0~.
    \end{aligned}
\end{equation}

In contrast to the other frame, the full Higgs branch in the tadpole frame contains several additional irreducible components besides the special Higgs branch and the twisted Higgs branch.
These extra components arise from the more intricate pattern of $J$- and $E$-term constraints specific to this frame.

Nevertheless, despite the presence of these additional branches, the Hilbert series of both the special Higgs branch and the twisted Higgs branch are identical to those obtained in the other frame.
This agreement provides a stringent check of frame independence for these distinguished components of the vacuum moduli space.

\subsubsection*{Example: \texorpdfstring{$(g_1,n)=(2,0)$}{(g1,n)=(2,0)} (tadpole frame)}\label{subsec:(2,0)frame1}

\begin{figure}[ht]
    \centering
    \includegraphics[width=0.9\linewidth]{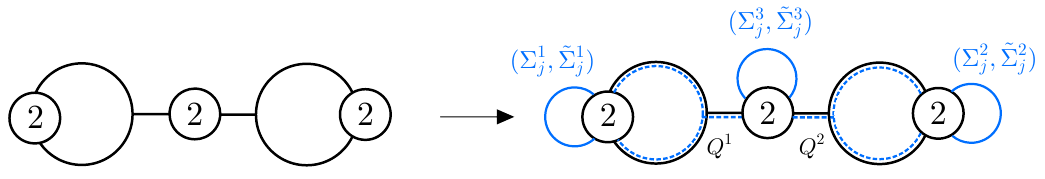}
    \caption{The quiver diagrams of the class $\mathcal{S}$ theory $\mathcal{T}[C_{2,0}]$ and the corresponding 2d (0,4) theory upon the reduction on $C_{g_2}$, in the ``dumbbell'' frame. For the 2d theory, the solid black lines represent two fundamental hypermultiplets $Q^I_{abc},I=1,2$. The solid blue line represents $3g_2$ adjoint twisted hypermultiplets $\Sigma_j^{J},\tilde\Sigma_j^{J},J=1,2,3$ and the dashed blue lines represent $2g_2$ fundamental Fermi multiplets $\Gamma^I_{jabc},~I=1,2,~j=1,\cdots g_2$.}
    \label{fig:reduce_g2n0_dumbbell}
\end{figure}

In the tadpole frame, the quiver for the case $(g_1,n) = (2,0)$ is shown in
Figure~\ref{fig:reduce_g2n0_dumbbell}. 
The corresponding $J$-term and $E$-term equations are
\begin{equation}\label{JE(2,0)}
    \begin{aligned}
    q^I_{abc}q^I_{\ap\bp\cp}\left((e_A)^{a\ap}\epsilon^{b\bp}\epsilon^{c\cp}+\epsilon^{a\ap}(e_A)^{b\bp}\epsilon^{c\cp}\right)&=0~,\\
        \left(q^1_{abc}q^1_{\ap\bp\cp}+q^2_{abc}q^2_{\ap\bp\cp}\right)\epsilon^{a\ap}\epsilon^{b\bp}(e_A)^{c\cp}&=0~,\\
        q^I_{\ap\bp\cp}\left(\sigma^{Ia\ap}_j\epsilon^{b\bp}\epsilon^{c\cp}+\epsilon^{a\ap}\sigma^{Ib\bp}_j\epsilon^{c\cp}+\epsilon^{a\ap}\epsilon^{b\bp}\sigma^{3c\cp}_j\right)&=0~,\\
        q^I_{\ap\bp\cp}\left(\tilde\sigma^{Ia\ap}_j\epsilon^{b\bp}\epsilon^{c\cp}+\epsilon^{a\ap}\tilde\sigma^{Ib\bp}_j\epsilon^{c\cp}+\epsilon^{a\ap}\epsilon^{b\bp}\tilde\sigma^{3c\cp}_j\right)&=0~,\\
        \sum_{j=1}^{g_2}\left[\sigma^J_j,\tilde\sigma^J_j\right]&=0~.
    \end{aligned}
\end{equation}

From the minimal prime decomposition of the ideal generated by \eqref{JE(2,0)}, we find that the full Higgs branch contains a special Higgs branch, a twisted Higgs branch, and several additional irreducible components in this frame.
\begin{itemize}
    \item \textbf{Special Higgs branch.}  
    The special Higgs branch has quaternionic dimension $2 g_2 + 1$ and is characterized by the constraints
    \begin{equation}
        \sigma^3_j = \tilde{\sigma}^3_j = 0~, 
        \qquad
        [\sigma_i^J , \sigma_j^J]
        = [\sigma_i^J , \tilde{\sigma}_j^J]
        = [\tilde{\sigma}_i^J , \tilde{\sigma}_j^J]
        = 0~,
    \end{equation}
    for all $i,j = 1, \ldots, g_2$ and $J = 1,2$.

    In this tadpole frame, the scalars in the twisted hypermultiplet
    $(\Sigma^3, \tilde{\Sigma}^3)$, which are \emph{not} associated with loop nodes in the quiver, vanish on the special Higgs branch, in agreement with the general constraint \eqref{constr_special_higgs_2}.
    In contrast, the scalars in the twisted hypermultiplets
    $(\Sigma^J, \tilde{\Sigma}^J)$ with $J=1,2$, which are associated with loop nodes, are allowed to acquire non-zero vacuum expectation values, as described in \eqref{constr_special_higgs_1}.

    The corresponding Hilbert series takes the form
    \begin{equation}
        G_{H,(2,0)}(t; g_2)
        =
        \frac{1 + \cdots}
        {(1 - t^2)^{4 g_2 + 1} (1 - t^4)}~.
    \end{equation}
    The explicit expressions for $g_2 = 0, \ldots, 3$ are listed in Table~\ref{tab:HS_(2,0)}.

A notable feature of these results is that, for $g_2 > 0$, the numerator of the Hilbert series is \emph{not} palindromic.
In particular, for $g_2 = 1$, we have computed the Hilbert series using two independent methods: the minimal prime decomposition of the ideal and the gluing method.
We find an agreement between the two approaches.

At present, however, the geometric interpretation underlying these non-palindromic Hilbert series remains unclear.
Understanding the geometry of these Higgs branches and the physical significance of this feature is an interesting open problem, which we leave for future investigation.

\begin{table}[ht]
\centering    \renewcommand{\arraystretch}{1.3}
\begin{tabular}{c|c|c}
\hline
{$g_2$} & {dim} & {Hilbert series}  \\
\hline
0 & 1 & $\frac{1+t^4}{\left(1-t^2\right) \left(1-t^4\right)}$ \\
\hline
1 & 3 & $\frac{1+2 t^2+8 t^3+6 t^4+8 t^5+10 t^6-3 t^8}{\left(1-t^2\right)^5 \left(1-t^4\right)}$ \\
\hline
2 & 5 & $\frac{1+12 t^2+16 t^3+55 t^4+112 t^5+88 t^6+112 t^7+103 t^8+16 t^9+12 t^{10}-15 t^{12}}{\left(1-t^2\right)^9 \left(1-t^4\right)}$  \\
\hline
3 & 7 & $\frac{1+30 t^2+24 t^3+292 t^4+440 t^5+794 t^6+1584 t^7+1426 t^8+1584 t^9+1426 t^{10}+440 t^{11}+300 t^{12}+24 t^{13}-138 t^{14}-35 t^{16}}{\left(1-t^2\right)^{13} \left(1-t^4\right)}$   \\
\hline
\end{tabular}
\caption{The quaternionic dimensions and unrefined Hilbert series of the special Higgs branches of the case $(g_1,n)=(2,0)$. In particular, the numerators are not palindromic for $g_2\geq 1$.}
\label{tab:HS_(2,0)}
\end{table}

\item  The Hilbert series of the twisted Higgs branch is given by the general result \eqref{eq:tw_HS_general} at $N_v=3$.
\end{itemize}

For the case $(g_1,n) = (2,0)$, there exists another duality frame whose quiver takes the form of a theta graph, referred to as the ``Yin-Yang'' quiver in~\cite{Hanany:2010qu}. 
However, with the limitations of standard desktop implementations of \texttt{Macaulay2}, we are unable to compute the corresponding Hilbert series in this alternative frame.

\section{Dimensional reduction of a single M5-brane}\label{app:single_M5}

In this appendix, we study the two-step dimensional reduction of the worldvolume theory of a single M5-brane on the product manifold
$C_{g_1,n} \times C_{g_2}$, with particular emphasis on the resulting zero modes.
The reduction proceeds in two stages, each accompanied by an appropriate partial topological twist to preserve the desired amount of supersymmetry.

We first reduce the 6d theory on $C_{g_1,n}$, obtaining a 4d $\mathcal{N}=2$ theory.
Subsequently, we further compactify this 4d theory on $C_{g_2}$, which yields a 2d $\mathcal{N}=(0,4)$ theory.
Since the worldvolume theory of a single M5-brane is free, the resulting 2d theory is also free.
As a consequence, the central charge can be determined directly by counting the physical degrees of freedom.
Below, we identify the field content through dimensional reduction and then compute the 2d central charge via this field-counting approach.

\medskip

The worldvolume theory of a single M5-brane is a 6d $\mathcal{N}=(2,0)$ free theory consisting of a single tensor multiplet.
Its field content includes a self-dual two-form $B_{MN}$, five real scalars $\Phi_{I}$ ($I=1,\ldots,5$), and four symplectic Majorana--Weyl fermions $\Psi_{\alpha}$ ($\alpha=1,2,3,4$).
Under the $SO(5)_R$ R-symmetry, these fields transform in the representations $\mathbf{1}$, $\mathbf{5}$, and $\mathbf{4}$, respectively.

Placing the theory on the spacetime
$\mathcal{M}_2 \times C_{g_1,n} \times C_{g_2}$, the 6d Euclidean Lorentz group decomposes as
\begin{equation}
SO(6)_E
~\longrightarrow~
SO(2)_{\mathcal{M}_2}
\times U(1)_{C_{g_1,n}}
\times U(1)_{C_{g_2}} \, .
\end{equation}
Punctures on the Riemann surfaces correspond to codimension-two defects in the 6d theory.
These defects give rise to flavor symmetries and matter couplings in the lower-dimensional theories obtained after compactification.

\paragraph{Reduction on $C_{g_1,n}$.}
The R-symmetry of the 6d $\mathcal{N}=(2,0)$ theory is $SO(5)_R \supset U(1)_r \times SU(2)_R$.
We perform a topological twist along $U(1)_{C_{g_1,n}}$ with $U(1)_r$,
\begin{equation}
U(1)_{C_{g_1,n}}^{\text{tw}}
\;=\;
\mathrm{diag}\,\bigl(U(1)_{C_{g_1,n}} \times U(1)_r\bigr)~.
\end{equation}
to preserve eight supercharges after the dimensional reduction on $C_{g_1,n}$~\cite{Gaiotto:2009we}.
The resulting theory is a 4d $\mathcal{N}=2$ theory on $\mathcal{M}_2\times C_{g_2}$ with R-symmetry $SU(2)_R\times U(1)_r$.

In what follows, we determine the 4d field content by reducing the 6d fields on the closed Riemann surface $C_{g_1}$.
The effects of punctures are incorporated separately through appropriate boundary conditions associated with codimension-two defects.


\begin{itemize}

  \item \textbf{Two-form field.}  
    The two-form field $B_{MN}$ transforms in the $\mathbf{15}$ of $SO(6)_E$ and is a singlet under $SO(5)_R$.
    Its field strength $H = \dd B$ admits the following decomposition upon reduction on $C_{g_1}$:
    \begin{equation}\label{H}
        H
        =
        \dd b
        + \sum_{k=1}^{g_1}
        \bigl(
        \dd a_k \wedge \omega_A^k
        + \dd \tilde{a}_k \wedge \omega_B^k
        \bigr)
        + \dd \varphi_0 \wedge \Omega~,
    \end{equation}
    where $\Omega$ is the volume form on $C_{g_1}$ and
    $\{\omega_A^k , \omega_B^k\}_{k=1,\ldots,g_1}$ form a basis of
    $H^1(C_{g_1},\mathbb{R})$.
    As a result, the 4d theory contains a two-form $b$, a real scalar $\varphi_0$, and $g_1$ Abelian gauge fields $a_k$.~\footnote{The fields $\tilde{a}_k$ are magnetic duals of $a_k$ and will not be treated as independent degrees of freedom in the following.}

   \item \textbf{Scalars.}  
    The five real scalars $\Phi_{I=1,\ldots,5}$ transform in the $\mathbf{5}$ of $SO(5)_R$.
    Under the decomposition
\begin{align}
SO(6)_E   \times SO(5)_R 
\quad  \to &   \quad 
SU(2)_R  \times SU(2)_l \times SU(2)_r \times  U(1)^{\text{tw}}_{C_{g_1}} \times U(1)_r
\nonumber \\
(\bf{1},{\bf 5})
\quad  \to & \quad 
\bf{(3, 1,1)}_{0,0} + \bf{(1,1,1)}_{\pm 1,\pm 1}
\end{align}
 where $SU(2)_\ell \times SU(2)_r$ is the holonomy group of the uncompactified four-manifold.
The first representation gives rise to three real 4d scalars $\varphi_{1,2,3}$.
The second representation carries charge $\pm1$ under $U(1)^{\text{tw}}_{C_{g_1}}$ and is therefore naturally interpreted as a one-form on $C_{g_1}$.
Expanding these twisted scalars in the harmonic basis
$\{\omega_A^k , \omega_B^k\}_{k=1,\ldots,g_1}$ of
$H^1(C_{g_1},\mathbb{R})$, in complete analogy with~\eqref{H},
produces $2g_1$ real zero modes, which pair up into $g_1$ complex scalars $\phi_k$ in 4d.

\item \textbf{Fermions.}  
    The 6d fermions $\Psi_{\alpha=1,\ldots,4}$ transform as $(\mathbf{4}^+,\mathbf{4})$ under $SO(6)_E \times SO(5)_R$.
    Their decomposition under the twisted symmetry group is
    \begin{align}
SO(6)   \times SO(5)_R 
\quad  \to &   \quad  SU(2)_R  \times SU(2)_l \times SU(2)_r \times   U(1)^{\text{tw}}_{C_{g_1}} \times U(1)_r \nonumber \\
 (\bf{4^+},{\bf 4})
\quad  \to &  \quad \bf{(2,2,1)}_{1,\frac{1}{2}}+\bf{(2,2,1)}_{0,-\frac{1}{2}}+\bf{(2,1,2)}_{0,\frac{1}{2}}+\bf{(2,1,2)}_{-1,-\frac{1}{2}}
\end{align}
The middle two representations yield one pair of left- and right-moving fermions $(\psi,\tilde{\psi})$ in 4d.
By contrast, the first and last representations carry non-zero charge under
$U(1)^{\text{tw}}_{C_{g_1}}$ and are therefore treated as one-forms on $C_{g_1}$.
Expanding them in the harmonic basis
$\{\omega_A^k , \omega_B^k\}_{k=1,\ldots,g_1}$ of
$H^1(C_{g_1},\mathbb{R})$
gives rise to $g_1$ additional pairs of left- and right-moving fermions
$(\lambda_k,\tilde{\lambda}_k)$, with $k=1,\ldots,g_1$.

\item  \textbf{Punctures.}  
The contribution of punctures on $C_{g_1}$ can be analyzed by reducing the Abelian 6d $\cN=(2,0)$ theory on the half-cylinder $S^1 \times \bR^+$. Reduction on the circle yields an Abelian 5d $\cN=2$ SYM theory, with the half-BPS boundary conditions of the 5d theory incorporating the puncture contributions.
Equivalently, after further reduction on an additional circle, specifying such a boundary condition translates to choosing one for a 4d $\cN=4$ SYM theory. The half-BPS boundary conditions of 4d $\cN=4$ SYM theory have been classified in~\cite{Gaiotto:2008sa}. For the $G=U(1)$ case, there are two basic types: the Dirichlet condition, which gives a free 3d $\cN=4$ (twisted) hypermultiplet, and the Neumann condition, which gives a free vector multiplet~\cite{Gaiotto:2008ak,Benini:2010uu}. We choose the Dirichlet condition for all punctures; when uplifted to 4d, each such puncture contributes one free $\cN=2$ hypermultiplet, resulting in a total of $n$ free hypermultiplets. 

\end{itemize}

These fields can be organized as 4d $\cN=2$ supermultiplets. In total, we have 
\begin{itemize}
    \item $g_1$ $\cN=2$ vector multiplets $V_{\text{4d},k}=(a_{k},\phi_{k}, \lambda_{k},\tilde{\lambda}_{k})$ with $k=1,2,\ldots,g_1$
    \item $(n+1)$ $\cN=2$ hypermultiplet $H_{\text{4d},i} (q,\tilde{q},\psi,\tilde{\psi})$ with $i=0,1,\ldots, n$.   In particular, the hypermultiplet $H_{\text{4d},0}$ arises from the scalar sector of the 6d tensor multiplet.
    The real scalar $\varphi_0$, which is a singlet under $SU(2)_R$, together with the three real scalars $\varphi_{1,2,3}$, which form a triplet of $SU(2)_R$, combine into two complex scalars transforming in the $\mathbf{2}$ of $SU(2)_R$, denoted by $(q,\tilde{q})$.
\end{itemize}

\paragraph{Reduction on $C_{g_2}$.}
We now reduce these free 4d $\mathcal{N}=2$ multiplets on $C_{g_2}$.
To preserve four supercharges, we perform a topological twist using the $U(1)_r$ symmetry, as explained in Section~\ref{sec:4dto2d}.
Upon dimensional reduction on $C_{g_2}$~\cite{Kapustin:2006hi,Putrov:2015jpa}, the resulting 2d (0,4) theory is again free and contains the following field content:
\begin{itemize}
    \item $g_1$ vector multiplets and $g_1 g_2$ twisted hypermultiplets,
    \item $n+1$ hypermultiplets and $(n+1) g_2$ Fermi multiplets.
\end{itemize}
The $g_1$ Abelian gauge fields are gapped in the infrared. Consequently, the 2d (0,4) theory in the infrared consists solely of free scalars and fermions.
The contribution of a real scalar to the central charge is $1$, while that of a Weyl fermion is $1/2$.
Accordingly, a (0,4) hypermultiplet or twisted hypermultiplet contributes $(c_L,c_R)=(4,6)$, whereas a (0,4) vector multiplet or Fermi multiplet contributes $(c_L,c_R)=(2,0)$.
From the above field content, we find that the central charges of a single M5-brane compactified on $C_{g_1,n} \times C_{g_2}$ are
\begin{align}
    c_R &= 6 \bigl( g_1 g_2 + n + 1 \bigr)~, \cr
    c_L &= 4 \bigl( g_1 g_2 + n + 1 \bigr)
    + 2 \bigl( g_1 + g_2 + n g_2 \bigr)~.
\end{align}
Both central charges are positive for arbitrary $g_1, g_2 > 0$.

\bibliographystyle{JHEP}
\bibliography{references}
\end{document}